\documentclass{jfm}

\usepackage{upgreek}
\usepackage{amsmath}
\usepackage{color}

\ifCUPmtlplainloaded \else
  \checkfont{msam10}
  \iffontfound
    \IfFileExists{amssymb.sty}
      {\typeout{^^JFound AMS Symbol fonts on the system, using the
                'amssymb' package.^^J}%
       \usepackage{amssymb}%
         \let\leq=\leqslant
         \let\geq=\geqslant
      }{}
  \fi
\fi

\newcommand{\drawline}[2]{\raisebox{2.5pt}{\vbox{\hrule width #1 pt height #2 pt}}}
\newcommand{\spacce}[1]{\hspace{#1pt}}
\newcommand{\solid}{\nobreak\mbox{\drawline{24}{0.5}\spacce{2}}}

\newcommand{\ccline}{\drawline{5.5}{0.5}\spacce{2}\drawline{1}{0.5}\spacce{2}}
\newcommand{\dline}{\drawline{5.5}{0.5}\spacce{2}}
\newcommand{\dashdot}{\nobreak\mbox{\ccline\ccline\dline}}

\title{Streak instability in turbulent channel flow: the seeding mechanism of large-scale motions} 
\shortauthor{M. de Giovanetti, H. J. Sung and Y. Hwang} 

\shorttitle{Streak instability: the seeding mechanism of large-scale motions}

\author
 {
 Matteo de Giovanetti\aff{1}\corresp{\email{m.de-giovanetti14@imperial.ac.uk}},
 Hyung Jin Sung\aff{2} and Yongyun Hwang\aff{1}
  }

\affiliation
{
\aff{1}
Department of Aeronautics, Imperial College London, \\South Kensington SW7 2AZ, UK
\aff{2}
Department of Mechanical Engineering, KAIST, 291 Daehak-ro,\\ Yuseong-gu, Daejeon 34141, South Korea}

\begin{document}

\maketitle

\begin{abstract}
It has often been proposed that the formation of large-scale motion (or bulges) is a consequence of successive mergers and/or growth of near-wall hairpin vortices. In the present study, we report our direct observation that large-scale motion is generated by an instability of an `amplified' streaky motion in the outer region (i.e. very-large-scale motion). We design a numerical experiment in turbulent channel flow up to $Re_\tau\simeq 2000$ where a streamwise-uniform streaky motion is artificially driven by body forcing in the outer region computed from the previous linear theory (Hwang \& Cossu, \emph{J. Fluid Mech.}, vol. 664, 2015, pp. 51--73). As the forcing amplitude is increased, it is found that an energetic streamwise vortical structure emerges at a streamwise wavelength of $\lambda_x/h\simeq 1-5$ ($h$ is the half-height of the channel). The application of dynamic mode decomposition and the examination of turbulence statistics reveal that this structure is a consequence of the sinuous-mode instability of the streak, a sub-process of the self-sustaining mechanism of the large-scale outer structures. It is also found that the statistical features of the vortical structure are remarkably similar to those of the large-scale motion in the outer region. Finally, it is proposed that the largest streamwise length of the streak instability determines the streamwise length scale of very-large-scale motion.
\end{abstract}

\section{Introduction}
Understanding the nature of coherent structures in wall-bounded turbulent shear flows has been a central research issue for many years. The near-wall region is very well understood, and contains two predominant structures: streaks \citep{Kline1967,Smith1983} and quasi-streamwise vortices \citep*{Kim1987,Jeong1997}. The interplay between the two structures comprises the so-called self-sustaining process, that lies in the heart of near-wall turbulence production \citep{Waleffe1997,Jimenez1999,Kim01}. An important sub-process of the self-sustaining process is the `lift-up' effect, by which the streamwise vortices transfer energy of the mean shear to the streaks. \citep*{Kim1971,Blackwelder1979,Landahl1980}. The lift-up effect is associated with the non-normality of the linearised Navier-Stokes operator, and it is understood to play a central role in determining the spanwise spacing of the near-wall streaks \citep{Butler1993,Chernyshenko2005,delAlamo2006a,Pujals2009,Cossu2009,Hwang2010a,Willis2010}. The streaks, amplified by the lift-up effect, subsequently undergo a secondary instability and/or transient growth that involves a rapid streamwise meandering motion (i.e. sinuous mode) \citep*{Hamilton1995,Schoppa2002,Cassinelli2017}. This is followed by the generation of vortical perturbations, which eventually develop into new streamwise vortices via non-linear mechanisms \citep{Hamilton1995,Schoppa2002}.

For the last two decades, important progress in the understanding of coherent structures in the logarithmic and outer regions has also been made. In particular, recent efforts have revealed that the `attached eddy hypothesis' \citep{Townsend1961,Townsend1976,Perry1982} is the key framework for the description of the coherent structures in the logarithmic and outer regions: for instance, the logarithmic wall-normal dependence of the turbulence intensities of wall-parallel velocity components \citep{Townsend1976,Perry1982,Perry1995,Jimenez2008,Marusic2013}, the emergence of $k_x^{-1}$ spectra ($k_x$ is the streamwise wavenumber) \citep{Perry1982,Perry1986,Nickels2005}, the linear growth of the spanwise integral length scale with distance from the wall \citep{Tomkins2003,Hwang2010d,Hwang2015} and the statistical and dynamical self-similarity of coherent structures \citep{delAlamo2006,Hwang2011,Hwang2015,Hellstrom2016,Hwang2016}. These findings directly support the notion that the entire logarithmic layer consists of a hierarchical form of self-similar coherent structures, as proposed in the seminal work of \cite{Townsend1961,Townsend1976} and \cite{Perry1982}.

A growing body of evidence further indicates that the coherent structures at a given length scale in the logarithmic and outer regions have a self-sustaining mechanism, essentially independent of those at other length scales \citep{Hwang2010d,Hwang2011,Hwang2015,Hwang2016}. Each of the self-sustaining coherent structures in the logarithmic region was found to consist of an elongated streak and streamwise vortical structures (or vortex packets), similar to that in the near-wall region \citep{Hwang2015}. In the outer region, structures of a similar form also emerge \citep{Hwang2015}, and it was proposed that the streak and the vortical structure correspond to very-large-scale motion (VLSM) \citep{Kim1999,delAlamo2003,Hutchins2007} and large-scale motion (LSM) \citep{Kovasznay1970}, respectively. The self-sustaining mechanism of the structures in the logarithmic and outer regions was also found to exhibit some important similarities to that in the near-wall region \citep{Flores2010,Hwang2016}. In particular, the generation of elongated streaks was well explained by the lift-up effect \citep{Cossu2009,Pujals2009,Hwang2010b,Hwang2010a,Willis2010}, and it was also shown that artificial inhibition of the lift-up effect destroys the self-sustaining mechanism \citep{Hwang2016}.

Despite this recent progress, the mechanism, by which the vortical structures in the logarithmic and outer regions are generated, remains an important issue of debate. Earlier studies proposed that mergers and/or growth of the hairpin vortices from the near-wall region could be the generation mechanism of the vortical structures in the logarithmic and outer regions \citep{Perry1986,Zhou01,Adrian2001}. The largest vortical structure generated by this process was thought to be the LSM \citep{Adrian2007} and a VLSM was subsequently interpreted as a concatenation of adjacent LSMs \citep*{Baltzer2013}. However, it should be pointed out that LSMs and VLSMs emerge even in the absence of motions in the near-wall and logarithmic regions \citep{Hwang2010d,Hwang2015}. They also remain virtually unchanged in the presence of significant disruption of the near-wall region \citep{Flores2006,Hutchins2007}. Finally, the scenario based on the mergers/growth of hairpin vortices does not provide any explanation as to why the streamwise length scale of LSMs is determined to be $\lambda_x\simeq 1-5\delta$ ($\lambda_x$ is the streamwise wavelength, and $\delta$ is the outer length scale, such as the boundary-layer thickness, half-height of the channel, and radius of the pipe).

Unlike VLSMs which predominantly carry streamwise turbulent kinetic energy, LSMs are the carriers of intense wall-normal and spanwise turbulent kinetic energy \citep{Hwang2015,Hwang2016}. Their streamwise extent is only $O(\delta)$, considerably smaller than the $O(10\delta)$ extent of VLSMs. These features are rather similar to the near-wall streamwise vortices, the streamwise size of which is only $\lambda_x^+\simeq 200-300$, compared to $\lambda_x^+\simeq 1000$ for near-wall streaks \cite[]{Jeong1997,Hwang2013}.This structural similarity suggests that the initiating mechanism of LSMs may be an instability of the amplified outer streak (VLSM), like that of streamwise vortices in the near-wall region. Indeed, a recent stability analysis with an eddy viscosity model revealed that the amplified streak becomes unstable at the typical streamwise length scale of LSMs and that the corresponding instability mode is characterised by a meandering motion of the streak (i.e. sinuous-mode instability) \citep{Park2011,Alizard2015}.

Despite this theoretical finding, the issue of how the vortical structures in the logarithmic and outer regions are formed remains unsettled, even among the work supporting the self-sustaining process. Indeed, \cite{Jimenez2013a,Jimenez2013} recently proposed that the Orr mechanism, which describes the transient amplification of a spanwise vortical perturbation by the mean shear, plays a role in the generation of the vortical structures. However, it should be pointed out that the Orr mechanism does not provide any description of the streamwise length-scale selection of the vortical structures, casting doubt on whether the Orr mechanism is the primary mechanism of the generation of vortical-structure. The issue of the streamwise length scale selection also appears in the recent analysis by \cite{Sharma2013}, who investigated triadically consistent interactions between the resolvent modes. In their analysis, the length scales (wavenumbers) of the modes involved in the triadic interaction had to be `chosen' from an observation of the existing database, but their analysis does not show why the system has to choose such a specific triadic interaction among many other possible combinations of the wave triad. Finally, it should be mentioned that most of the previous numerical studies have investigated the dynamics of the self-sustaining process only with a very short streamwise computational domain \citep{Flores2010,Hwang2016}. This restriction does not allow one to explore the mechanism by which the streamwise length scale of the coherent structures at a given spanwise length scale is determined.

\begin{figure}
  \centerline{\includegraphics[width=0.90\textwidth]{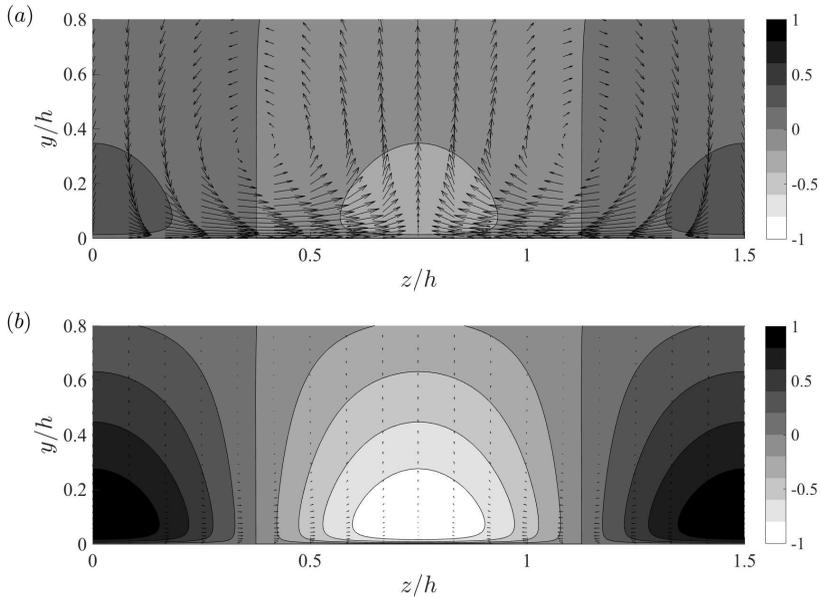}}
  \caption{Cross-streamwise view of $(a)$ the body forcing $\mathbf{\tilde{f}}(y,z)(=\mathbf{f})$ and $(b)$ the response $\mathbf{\tilde{u}}(y,z)$ with (\ref{eq:2.2}). The contours indicate the streamwise component, while the vectors indicate the wall-normal and spanwise components. Here, the contour levels in $(a)$ are normalised by the maximum value of the wall-normal component to highlight that the forcing is dominated by the cross-streamwise components (i.e. streamwise vortices), whereas those in $(b)$ are normalised by the maximum value of the streamwise component to show that the response is dominated by the streamwise component (streak). The contours and the vectors in $(a)$ and $(b)$ are plotted with the same scale.}
\label{forc_profile}
\end{figure}

Given the ongoing discussion about the formation of the vortical structures in the logarithmic and outer regions, the existence of streak instability and the related streamwise length-scale selection mechanism in `real' turbulent flows must be examined. In particular, this is the only scenario that provides an explanation of the streamwise length-scale selection of the vortical structures. The goal of the present study is therefore to demonstrate that the streak instability is the seeding mechanism of the vortical structure in the outer region (i.e. LSM). To this end, we design a numerical experiment that enables us to access the streak instability in a `real' turbulent flow. We introduce a body forcing, computed with the previous linear theory \cite[]{Hwang2010a}, to drive long streaks in the outer region (i.e. an idealised form of VLSMs). The forcing amplitude is then gradually elevated to examine the emergence of streak instability and the corresponding eigenstructure. To carefully detect the eigenstructure, a dynamic mode decomposition (DMD) \citep{Schmid2010,Jovanovic2014} is subsequently employed. From the designed numerical experiment, we will indeed see that sinuous-mode streak instability and the related vortical structure emerge with a streamwise length scale of $\lambda_x\simeq 1-5\delta$.

\section{Numerical experiment}\label{sec:NumExp}

The numerical experiment in the present study has been performed in turbulent channel flow using a near-wall-resolved large-eddy simulation (LES). The streamwise, wall-normal and spanwise directions are denoted as \textit{x}, \textit{y} and \textit{z}, respectively, and the corresponding velocities are defined as \textit{u}, \textit{v} and \textit{w}, respectively. The lower and upper walls of the channel are set at $y=0$ and $y=2h$, respectively where $h$ denotes the half-height of the channel. The numerical solver discretises the Navier-Stokes equation in the streamwise and spanwise directions using Fourier series with the 2/3 dealiasing rule, whereas the wall-normal direction is discretised using second-order central differences. The time-stepping is performed semi-implicitly with a fractional step method: the Crank-Nicolson method is used for the terms with wall-normal derivatives and a third-order Runge-Kutta method is used for all the remaining terms. The residual stress in the present LES is approximated by an eddy-viscosity-based model proposed by \cite{Vreman01}. The present LES has been previously validated over a range of Reynolds numbers from $Re_\tau \simeq 1000$ to $Re_\tau \simeq 4000$, as detailed in \citet{deGiovanetti2016}. All the numerical simulations in the present study are performed by imposing constant mass flux across the channel.

\subsection{The body forcing}\label{ssec:forc}
Given the scope of the present study, the role of the body forcing is to drive a long streaky motion in the flow field. We obtain such a body forcing by following \cite{Hwang2010a}, who computed a forcing that drives streaks in an optimal manner using a linear theory. We note that since the forcing is merely to drive the streak, other forms of the forcing can also be considered as long as they can drive a streaky motion. This issue will be further discussed in \S\ref{sec:4.4}. In this section, we briefly summarise the procedure to compute the forcing, and the reader may refer to \cite{Hwang2010a} for further details.

We start by considering the triple decomposition of a given flow field as in \cite{Reynolds1972}:
\begin{equation}\label{eq:2.1}
\mathbf{u}=\mathbf{U}_0+\tilde{\mathbf{u}}+\mathbf{u}',
\end{equation}
where $\mathbf{u} = (u,v,w)$ is the velocity field, $\mathbf{U}_0 = (U_0(y),0,0)$ the mean velocity obtained by averaging in time and in two homogeneous directions, $\tilde{\mathbf{u}}$ the so-called `organised wave', and $\mathbf{u}'$ the random turbulent velocity fluctuation. If we strictly follow \cite{Reynolds1972}, $\mathbf{U}_0$ is the mean velocity in the absence of the forcing and $\tilde{\mathbf{u}}$ is the motion induced by a systematically controllable external forcing (e.g. vibrating ribbon). Therefore, $\mathbf{U}_0$ is obtained by an experiment or a numerical simulation without any external forcing, whereas $\tilde{\mathbf{u}}$ is obtained by applying a suitable ensemble (or phase) average to the flow fields in the presence of the given forcing. For this reason, in the absence of any external forcing, $\tilde{\mathbf{u}}$ should become zero, and, in such a circumstance, (\ref{eq:2.1}) becomes identical to the standard Reynolds decomposition.

Now, we consider a deterministic body forcing ${\mathbf{f}} = ({f}_u,{f}_v,{f}_w)$ that will be used to drive a streak in the numerical simulation to be performed. We set the mean of the forcing, obtained by averaging in time and in two homogenous directions, to be zero (see also (\ref{eq:2.4b})) so that the forcing can be captured completely by the chosen ensemble average, i.e. ${\mathbf{f}}=\tilde{\mathbf{f}}$. If the size of ${\mathbf{f}}$ is small, $\tilde{\mathbf{u}}$ also becomes small. In this case, the following linearised equation for $\tilde{\mathbf{u}}$ is obtained if an appropriate closure is provided to describe the effect of $\mathbf{u}'$ on the evolution of $\tilde{\mathbf{u}}$ \citep{Reynolds1972}:
\begin{subequations}\label{eq:2.2}
\begin{equation}
\nabla \cdot \tilde{\mathbf{u}} = 0,
\end{equation}
\begin{equation}
\frac{\partial \tilde{\mathbf{u}}}{\partial t} + (\mathbf{U}_0 \cdot \nabla )\tilde{\mathbf{u}} + (\tilde{\mathbf{u}} \cdot \nabla)\mathbf{U}_0 = -\frac{1}{\rho} \mathbf{\nabla} \tilde{p} + \mathbf{\nabla \cdot} [(\nu+\nu_t) (\nabla \tilde{\mathbf{u}} + \nabla \tilde{\mathbf{u}}^T)] + \tilde{\mathbf{f}},
\end{equation}
\end{subequations}
where $\tilde{p}$ is the pressure induced by the forcing and $\nu_t$ the eddy viscosity that models the effect of the surrounding turbulence $\mathbf{u}'$ on $\tilde{\mathbf{u}}$. As in previous studies \cite[e.g.][]{Reynolds1967}, the well-known Cess model is considered for $\nu_t$:
\begin{equation}\label{eq:2.3}
\nu_t (\eta) = \frac{\nu}{2} \Big\{1+ \frac{\kappa^2 Re_{\tau}^2}{9}(1- \eta^2 )^2 (1+2 \eta^2 )^2 \times {1- e^{{[(|\eta|-1) Re_{\tau}/A]}^{2}}} \Big\} ^{\frac{1}{2}} - \frac{\nu}{2},
\end{equation}
where $\eta=(y-1)/h$, $\kappa= 0.426$ and $A=25.4$ from \cite{delAlamo2006a}. The mean-velocity profile $U_0(y)$ is subsequently obtained by applying Prandtl's mixing length model to (\ref{eq:2.3}): $\nu_t dU_0/dy=-\overline{u'v'}$ where the overbar denotes average in time and homogeneous directions. For further details on the profiles of $\nu_t(y)$ and $U_0(y)$, the reader may refer to \cite{delAlamo2006a} and \cite{Pujals2009}.

The body forcing that yields the largest response of (\ref{eq:2.2}) has been previously computed in \cite{Hwang2010a}. This forcing is given in the form of a counter-rotating vortical motion and it creates a streak by optimally utilising the lift-up effect (see also figure \ref{forc_profile}). Here, we compute this forcing to drive a streak in the simulation. Given the spatial homogeneity of the mean flow and the eddy viscosity in the streamwise and the spanwise directions, the solution of (\ref{eq:2.2}) is written as
\begin{subequations}
\begin{equation}
\tilde{\mathbf{u}}(x,y,z,t)=\hat{\mathbf{u}}(y,t;k_x,k_z)e^{{i(k_x x+k_z z)}},
\end{equation}
where $k_x$ and $k_z$ are the streamwise and spanwise wavenumbers, respectively. The body forcing is also written in the form of
\begin{equation}\label{eq:2.4b}
\tilde{\mathbf{f}}(x,y,z,t)=\hat{\mathbf{f}}(y,t;k_x,k_z)e^{{i(k_x x+k_z z)}}.
\end{equation}
\end{subequations}
We now assume that the forcing is harmonic in time with a prescribed real frequency $\omega_f$, i.e. $\hat{\mathbf{f}}(y,t)=\hat{\mathbf{f}}_w(y)e^{i\omega_ft}$. Since (\ref{eq:2.2}) is a linear time-invariant system, its solution becomes $\hat{\mathbf{u}}(y,t)=\hat{\mathbf{u}}_w(y)e^{i\omega_ft}$ after initial transience. Then, the forcing profile leading to the largest response will be computed by solving the following optimisation problem:
\begin{equation}\label{eq:resp}
\max_{\omega_f}\max_{\|\hat{\mathbf{f}}_w\|\neq0} \frac{{||\hat{\mathbf{u}}_w||}} {{||\hat{\mathbf{f}}_w||}},
\end{equation}
where $||~\large{\cdot}~||^2\equiv(1/h)\int^{h}_{-h}(\cdot)^H(\cdot)~\mathrm{d}y$ and $(\cdot)^H$ indicates the complex conjugate transpose.

The solution procedure of (\ref{eq:resp}) typically involves the computation of the 2-norm of the so-called resolvent operator \citep{Schmid2001}. In this study, the optimisation problem (\ref{eq:resp}) is solved by repeating the calculation by \cite{Hwang2010a} using their numerical solver with the same wall-normal resolution. Since the present study is concerned with the instability of the streak at large scales (i.e. VLSM), the spanwise spacing of the forcing needs to be identical to that of VLSM. Since the mean spanwise spacing of the VLSM in turbulent channel flow was found to be $\lambda_z\simeq1.5h$ \cite[]{delAlamo2003,Hwang2015}, the spanwise wavenumber of the forcing is chosen to be $k_z =2\pi/1.5h$. To generate the streamwise-uniform streaky mean flow, the streamwise wavenumber and the forcing frequency are chosen to be $k_x=0$ and $\omega_f=0$, respectively. We note that, for a given spanwise wavenumber $k_z$, this choice of $k_x$ and $\omega_f$ yields the largest response of $\tilde{\mathbf{u}}$ with (\ref{eq:2.2}) (see below for further explanation). Also, the streamwise-uniform streaky mean flow allows us to examine all possible streamwise wavelengths of the streak instability.

Fig. \ref{forc_profile} shows the cross-streamwise view of the computed optimal forcing profile $\tilde{\mathbf{f}}$ and the corresponding response $\tilde{\mathbf{u}}$ of (\ref{eq:2.2}). The forcing $\tilde{\mathbf{f}}$ is given in the form of a pair of counter-rotating streamwise vortices, and is dominated by the cross-streamwise velocity components (figure \ref{forc_profile}$a$). On the other hand, the response velocity field $\tilde{\mathbf{u}}$ is dominated by a streaky motion of the streamwise velocity and its cross-streamwise velocity components are much weaker than the streamwise one (figure \ref{forc_profile}$b$). This indicates that the cross-streamwise components in the forcing are used to amplify the streamwise velocity streak in the response field. This is the so-called `lift-up' effect, in which the streamwise velocity component is highly amplified by the presence of a small amount of the cross-streamwise velocity components (the wall-normal one in particular).

The lift-up effect can be better understood from the wall-normal velocity and vorticity form of the linearised equation for $\hat{\mathbf{u}}(y,t;k_x,k_z)$: i.e.
\begin{subequations}\label{eq:OSSQ}
\begin{eqnarray}
\frac{\partial}{\partial t}
\left[\begin{array}{c}
   \hat{v}  \\
   \hat{\omega}_y  \\
\end{array}\right]
&=&\left[\begin{array}{cc}
   \Delta^{-1}\mathcal{L_{OS}} & 0 \\
   -ik_z U' & \mathcal{L_{SQ}}  \\
\end{array}\right]
\left[\begin{array}{c}
   \hat{v}  \\
   \hat{\omega}_y  \\
\end{array}\right] \\
&+&\left[\begin{array}{ccc}
   -ik_x \Delta^{-1}\mathcal{D} & - k^2 \Delta^{-1} & -ik_z\Delta^{-1} \mathcal{D} \\
   ik_z & 0 & -ik_x \\
\end{array}\right]
\left[\begin{array}{c}
   \hat{f}_x  \\
   \hat{f}_y  \\
   \hat{f}_z  \\
\end{array}\right], \nonumber
\end{eqnarray}
where
\begin{eqnarray}
\label{eq:OSSQop}
\mathcal{L_{OS}}&=&-ik_x(U \Delta-U'')
       +\nu_T \Delta^2+2 \nu_T' \Delta \mathcal{D} +\nu_T''(\mathcal{D}^2+k^2),\\
\mathcal{L_{SQ}}&=&-ik_x U + \nu_T \Delta+\nu_T' \mathcal{D}.
\end{eqnarray}
\end{subequations}
Here, $\hat{\omega}_y$ is the wall-normal vorticity of $\hat{\mathbf{u}}(y,t;k_x,k_z)$, $\nu_T=\nu+\nu_t$, $\Delta=\mathcal{D}^2-k^2$, $k^2=k_x^2+k_z^2$, and $\mathcal{D}$ and $'$ denote $\partial/\partial y$. In (\ref{eq:OSSQ}), it is important to note that the equation for $\hat{\omega}_y$ (the Squire equation) is passively coupled with the equation for $\hat{v}$ (the Orr-Sommefeld equation) via the off-diagonal term $-ik_z U'$. Therefore, the energy transfer between the wall-normal velocity and vorticity in (\ref{eq:OSSQ}) should always take place from $\hat{v}$ to $\hat{\omega}_y$ \citep{Schmid2001}. Furthermore, it should be noted that, for given $\hat{v}$, the amplification of $\hat{\omega}_y$ is solely governed by the Squire equation. Therefore, the largest amplification by the optimal forcing is expected in the set of streamwise and spanwise wavenumbers which would minimise the effect of dissipation in the Squire equation, while maximising the role of its driving term. In the Squire equation, the effect of dissipation is roughly proportional to $k^2(=k_x^2+k_z^2)$ , whereas the driving term is only proportional to $k_z$. Therefore, the largest possible amplification by the forcing is expected at $k_x=0$ and $k_z \simeq O(U'l/\nu_T)$ with $l$ being the length scale of $\omega_y$, as also confirmed by numerous previous studies \cite[e.g.][]{Schmid2001}. In this case, the main energy transfer between the velocity components of the linearised Navier-Stokes equation should be from $\hat{v}$ to $\hat{u}$. This is also why the optimal forcing is featured by large cross-streamwise components and its response is dominated by a large streamwise component (see figure \ref{forc_profile}).

Finally, it is important to mention that the linearised equation (\ref{eq:OSSQ}) does not allow for any mechanism of energy transfer from $\hat{\omega}_y$ (i.e. $\hat{u}$ and $\hat{w}$) to $\hat{v}$ because the equation for $\hat{v}$ is independent of $\hat{\omega}_y$. Such a mechanism is only achieved through the nonlinearity of the Navier-Stokes equation. Therefore, describing the dynamics of the wall-normal velocity with any kind of linear theory would be physically unmeaningful and incomplete. We will see that the streak instability mechanism provides a sound explanation for the generation of the wall-normal velocity from the highly amplified streamwise velocity structure (i.e. streak). In this respect, it is also worth mentioning that studying the streak instability mechanism would provide physical insight into design of constant or dynamical modifications of the linearized Navier-Stokes equations around turbulent mean velocity. Therefore, this effect may be modelled by various forms of deterministic and stochastic forcing introduced into the linearized dynamics, as also recently shown by \cite{Zare2017}.

\subsection{Dynamic mode decomposition}\label{ssec:DMD}
Dynamic mode decomposition (DMD) \cite[e.g.][]{Schmid2010} is employed to detect the eigenstructure of the streak instability using a set of flow-field snapshots detailed in $\S$\ref{ssec:DMDre}. The DMD is an increasingly popular post-processing technique for analysis of the temporal and spatial evolution of given flow structures of interest. It approximates the flow-field evolution with a finite-dimensional linear time-invariant (LTI) dynamical system. The DMD employed in the present study follows the one proposed by \citet*{Jovanovic2014} and, here, we only provide a brief summary of their approach.

We consider a sequence of equally time-spaced complex-valued snapshots with a time interval $\Delta t$, such that $\psi_n\equiv\psi(n\Delta t)$ where $\psi_n$ is the flow field at $t=n\Delta t$ for $n=0,1,2,...,N$. If the temporal evolution of the given flow field is assumed to be generated by a discrete-time LTI system, it can be written as
\begin{equation}\label{eq:2.6}
\psi_{n+1}=\mathbf{A}\psi_n.
\end{equation}
The snapshots, each composed of \textit{M} spatial points, are then collected in two ordered matrices $\boldsymbol{\Psi}_0$ and $\boldsymbol{\Psi}_1$, such that
\begin{subequations}
\begin{equation}
\boldsymbol{\Psi}_0=[\psi_0, \psi_1, \psi_2,  \dots, \psi_{N-1} ] ~~~ \in \mathbb{C}^{M \times N},
\end{equation}
\begin{equation}
\boldsymbol{\Psi}_1=[ \psi_1, \psi_2, \psi_3, \dots, \psi_{N} ] ~~~ \in \mathbb{C}^{M \times N}.
\end{equation}
\end{subequations}
The relation between $\boldsymbol{\Psi}_0$ and $\boldsymbol{\Psi}_1$ is subsequently given by
\begin{equation}
\boldsymbol{\Psi}_1=\mathbf{A}\boldsymbol{\Psi}_0.
\end{equation}
Here, the matrix $\mathbf{A}$ may be a suitable approximation of the so-called Koopman operator, especially when the snapshots are collected from a statistically stationary flow. A DMD is essentially an algorithm that computes eigenvalues and the corresponding eigenmodes (dynamic modes) of $\mathbf{A}$ only using the snapshots $\psi_n$.

The approximation of $\mathbf{A}$ in this study is made by combining with POD (proper orthogonal decomposition). This approach enables us to robustly implement the DMD, as it prevents mathematically singular construction of the companion matrix of  $\mathbf{A}$ from noise and other uncertainties in the snapshots (e.g. small-scale background turbulence in the present study) \citep{Schmid2010}. The POD is performed with an economy-size singular value decomposition of the snapshot matrix $\boldsymbol{\Psi}_0$, i.e. $\boldsymbol{\Psi}_0=\mathbf{U\Sigma V}^H$, where $\mathbf{\Sigma}=diag\{\sigma_1,\sigma_2,...,\sigma_{N_p}\}$ with non-zero real positive singular value $\sigma_m$ for $m=1,2,..,N_p$ ($N_p\leq N$), and $\mathbf{U} \in  \mathbb{C}^{M \times N_p}$ and $\mathbf{V} \in \mathbb{C}^{N \times N_p}$ are unitary matrices. Each POD mode is then given by the $m$-th column vector of $\mathbf{U}$ with the corresponding energy $\sigma_m$, and its temporal dynamics are described by the $m$-th column vector of $\mathbf{V}$. Here, if one is to use only the first $r$ POD modes ($r\leq N_p$), $\boldsymbol{\Psi}_0\simeq \mathbf{U}_r\mathbf{\Sigma}_r \mathbf{V}_r^H$ where $\mathbf{\Sigma}_r=diag\{\sigma_1,\sigma_2,...,\sigma_r\}$, $\mathbf{U}_r \in  \mathbb{C}^{M \times r}$ and $\mathbf{V}_r \in \mathbb{C}^{N\times r}$.

Projection of $\mathbf{A}$ onto the subspace of the first $r$ POD modes is then given by
\begin{equation}\label{DMD1}
\mathbf{{S}}=\mathbf{U}_r^H\mathbf{A}\mathbf{U}_r=\mathbf{U}_r^H\boldsymbol{\Psi}_1\mathbf{V}_r\mathbf{\Sigma}_r^{-1},
\end{equation}
where $\mathbf{{S}}\in \mathbb{C}^{r \times r}$. The eigenvalues of $\mathbf{{S}}$, denoted by $\mu_j$, then approximate some of the leading eigenvalues of $\mathbf{A}$ and the corresponding eigenvectors $\mathbf{y}_j$ construct the dynamic modes from
\begin{equation}
\phi_j=\mathbf{U}_r\mathbf{y}_j.
\end{equation}
Any snapshot may then be expressed in terms of a series expansion of the dynamic modes, such that:
\begin{equation}\label{eq:2.11}
\psi_n \simeq \sum_{j=1}^r \alpha_j \mu_j^n\phi_j,
\end{equation}
where $\alpha_j$ is the amplitude of each dynamic mode $\phi_j$, which represents its relative contribution. In this study, $\alpha_j$ has been calculated following the standard (non-sparsity-promoting) optimisation procedure in \citet{Jovanovic2014}.

\section{Results}\label{sec:results}

\begin{table}
  \begin{center}
\def~{\hphantom{0}}
 \begin{tabular}{c p{0.3cm} c c p{0.3cm} c  c p{0.2cm} c   p{0.3cm} c c c }
\setlength{\tabcolsep}{15pt}
 $Case$ &&$Re_m$& $Re_{\tau}$& &$L_x/h$&$L_z/h$ && $N_x \times N_y \times N_z$ & &$\Delta x^+$ & $\Delta z^+$ & $\Delta y^+_1$ \\ \\
 S490  &&  17800  & 494 && 25.0 & 3.0 && $288\times 81 \times 72$ && 64.7 & 32.3 & 1.45   \\
 S960  &&  38100  & 960  && 25.0 & 3.0 && $576\times 81 \times 144$  && 62.5 & 33.4 & 1.80         \\
 S2050 &&  89100  & 2056 && 12.0 & 3.0 && $576\times 129 \times 288$ && 64.3 & 32.1 & 1.44   \\
 \end{tabular}
  \caption{Simulation parameters of the unforced cases (before dealiasing).}
  \label{tab:DomainSummary}
  \end{center}
\end{table}

The designed experiment is carried out with a set of numerical simulations documented in table \ref{tab:DomainSummary}. Three different Reynolds numbers are considered, but most of the data presented in this section will be from the numerical simulations at $Re_\tau\simeq 960$ (S960). However, we note that the findings at this Reynolds number are qualitatively the same as those at the other two Reynolds numbers (see e.g. \S\ref{ssec:DMDre}). The body forcing computed in $\S$\ref{ssec:forc} is implemented in each of the numerical simulations and the forcing amplitude is gradually increased to drive stronger streaks. The forcing amplitude is defined as
\begin{equation}\label{eq:3.1}
||\mathbf{{f}}||=\sqrt{\frac{2}{V}\int_V \mathbf{{f}}^H\mathbf{{f}}~dV},
\end{equation}
where \textit{V} is the total volume of the computational domain. We note that the factor 2 is introduced here to make the norm in (\ref{eq:3.1}) identical to the one in (\ref{eq:resp}) and that $\mathbf{{f}}=\mathbf{\tilde{f}}$ by the definition of the forcing introduced in \S\ref{ssec:forc}.

\begin{figure}
  \centerline{\includegraphics[width=1.12\textwidth]{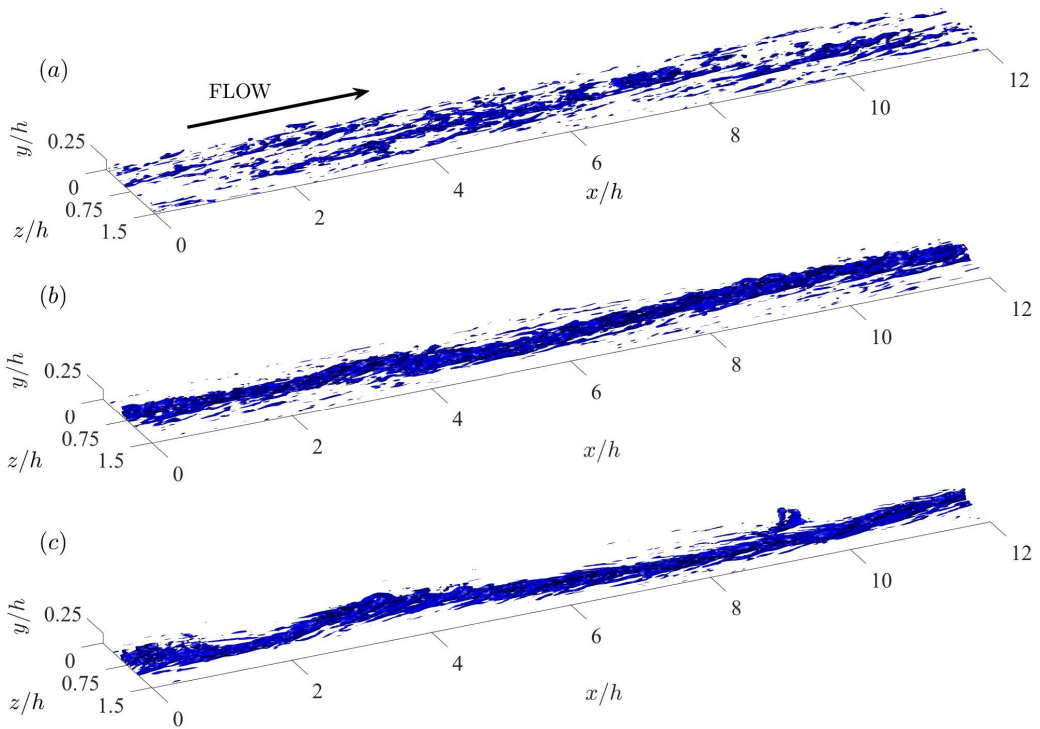}}\vspace*{-8mm}
  \caption{Iso-surfaces of instantaneous streamwise velocity (${u'}^+=-3.5$) for S960: (\textit{a}) $||\mathbf{{f}}h/u_{\tau,\mathrm{ref}}^2||=0.31$ ($A_s^+=1.44$), (\textit{b}) $||\mathbf{{f}}h/u_{\tau,\mathrm{ref}}^2||=0.94$ ($A_s^+=3.42$), (\textit{c}) $||\mathbf{{f}}h/u_{\tau,\mathrm{ref}}^2||=6.11$ ($A_s^+=5.30$).}
\label{Inststreak}
\end{figure}

Application of the forcing is found to affect friction drag and the related $Re_\tau$. However, for all forcing amplitudes considered, there was no case resulting in any drag reduction, consistent with the recent work by \citet{Canton2016}. For small forcing amplitudes, the friction Reynolds number $Re_\tau$ of the simulations changes very little. However, as the forcing amplitude is increased further, $Re_\tau$ increases significantly: see table \ref{tab:SimSummary} for the case of the S960 simulations. Here, it should be noted that the elevated skin friction is not due to any interaction with structures at smaller length scales in the near-wall and logarithmic regions. In fact, the introduction of the large-amplitude forcing has been found to destroy a significant amount of the structures in these regions (see figure \ref{spectra_comp}). Instead, the increase in skin friction for such large forcing amplitudes appears to be due to direct effect of the forcing on the wall shear stress. This interpretation is consistent with our recent findings \citep{Hwang2016,deGiovanetti2016} where the lift-up effect was found to be an important skin-friction generation mechanism of the structures in the logarithmic and outer regions. Indeed, the forcing in the present study, given in the form of counter-rotating vortices (figure \ref{forc_profile}), is designed to generate a streak by promoting the lift-up effect.

\begin{figure}
  \centerline{\includegraphics[width=1.05\textwidth]{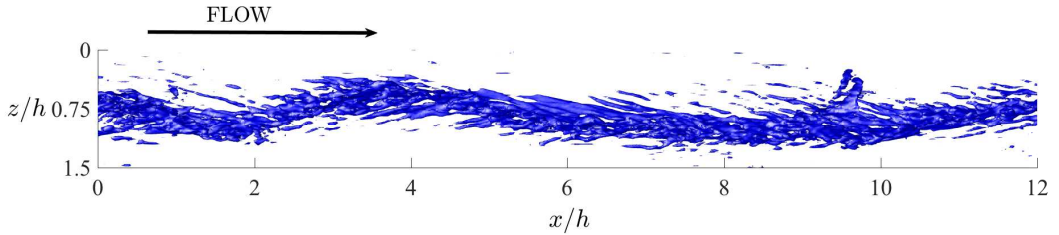}}
  \caption{Top view of the iso-surfaces of instantaneous streamwise velocity (${u'}^+=-3.5$) in figure \ref{Inststreak} $(c)$. Here, the flow direction is from left to right.}
\label{streak_top}
\end{figure}

\begin{table}
  \begin{center}
\def~{\hphantom{0}}
 \begin{tabular}{c c c  c  c  c  c  c  c  c }
 \setlength{\tabcolsep}{25pt}
  $\|\mathbf{{f}}h/u_{\tau,\mathrm{ref}}^2\|$ & 0 & 0.11 & 0.31 & 0.68 & 0.94 & 1.90 & 3.85 & 6.11 & 8.06 \\ [3pt]
   \hline
  $Re_\tau$ & ~960~ & ~967~ & ~992~ & ~1021~ & ~1029~ & ~1034~ & ~1042~ & ~1078~ & 1121 \\
  $\|\mathbf{\tilde{u}}_s^+\|$ & 0 & 0.63 & 1.90 & 2.53 & 2.76 & 3.00 & 3.42 & 3.79 & 4.04 \\
  $A_s^+$ & 0 & 0.54 & 1.44 & 2.96 & 3.42 & 4.04 & 4.73 & 5.30 & 5.72 \\
  \end{tabular}
  \caption{Summary of flow response with respect to the forcing amplitude (S960).}
  \label{tab:SimSummary}
  \end{center}
\end{table}

It should finally be mentioned that the main focus of the present study is the comparison of the reference and forced simulations in order to identify the flow structures associated with streak instability statistically and dynamically. Therefore, all data from the forced simulations are presented by scaling with the friction velocity of the (unforced) reference simulations, $u_{\tau,\mathrm{ref}}$.

\subsection{Streak amplification}\label{sec:3.1}

Figure \ref{Inststreak} reports the instantaneous streamwise velocity fluctuation of case S960 on increasing the forcing amplitude. Given the spanwise wavelength of the forcing ($\lambda_z/h=1.5$), two streaks are driven in our computational domain (see table \ref{tab:DomainSummary}). Here, for brevity, we only present half of the spanwise domain. As the forcing amplitude is gradually increased (from figure \ref{Inststreak}$a$ to $c$), a streaky structure covering the entire streamwise domain clearly emerges. It is also interesting to note that the highly amplified streak appears to meander along the streamwise direction (figure \ref{Inststreak}$c$) and this is clearly seen in figure \ref{streak_top}, where a top view of figure \ref{Inststreak} ($c$) is given. This meandering behaviour of the driven streak is reminiscent of that reported for VLSMs in a turbulent boundary layer \citep{Hutchins2007}. However, more quantitative discussion of this feature will be deferred to \S\ref{ssec:spec} and \S\ref{ssec:DMDre}.

The streaky structure driven by the body forcing in the numerical simulations is detected statistically using the definition of the triple decomposition in (\ref{eq:2.1}). Since the applied body forcing is steady and uniform in the streamwise direction, it is evident that the ensemble average in (\ref{eq:2.1}) is replaced by the average in time and the streamwise direction. The organised wave in a simulation with the forcing is then obtained by
\begin{equation}\label{eq:3.2}
\mathbf{\tilde{u}}_s=\langle\mathbf{u}-\mathbf{U}_0\rangle_{x,t},
\end{equation}
where the subscript `$s$' on the left-hand side denotes the organised wave component obtained by a numerical simulation and $\langle \cdot \rangle_{x,t}$ indicates the average in time and in the streamwise direction. We also recall that $\mathbf{U}_0=(U_0(y),0,0)$ is the mean velocity of the reference (unforced) simulation.

\begin{figure}
  \centerline{\includegraphics[width=1.15\textwidth]{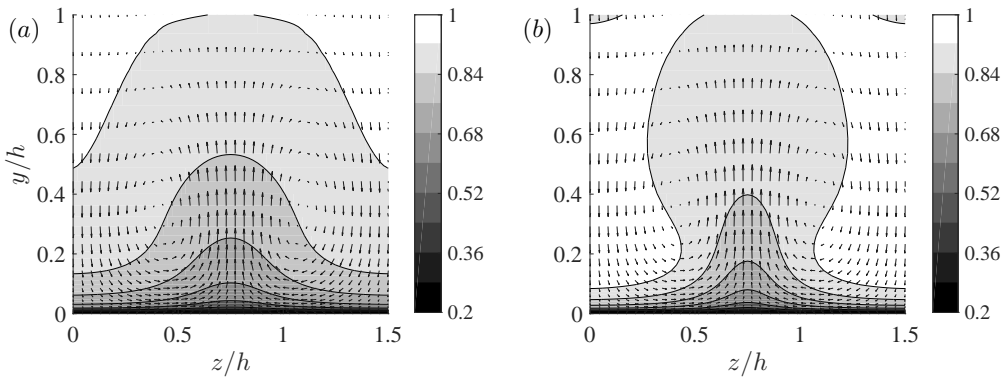}}
  \caption{Mean cross-stream flow field for S960: (\textit{a}) $\|\mathbf{{f}}h/u_{\tau,\mathrm{ref}}^2\|=0.31$ ($A^+_s=1.44$); (\textit{b}) $\|\mathbf{{f}}h/u_{\tau,\mathrm{ref}}^2\|=1.90$ ($A^+_s=4.04$). The contours indicate the streamwise velocity, and the arrows represent the wall-normal and spanwise velocities. The contours are normalised by the centreline velocity of each simulation.}
\label{Avfield}
\end{figure}

The cross-streamwise view of the mean velocity ($\langle\mathbf{u}\rangle_{x,t}=\mathbf{U}_0+\mathbf{\tilde{u}}_s$) is shown in figure \ref{Avfield} for two different forcing amplitudes. The appearance of a low-velocity region is evident around $z/h=0.75$, where the streamwise momentum is expected to be pumped away from the wall by the counter-rotating vortical forcing (see also figure \ref{forc_profile}). The low-speed region extends along the entire wall-normal location including the logarithmic layer ($0.05 <y/h< 0.2$ at this Reynolds number). Two symmetric high-velocity regions arise both sides of the low-velocity region, where the streamwise momentum is supposed to be brought down to the wall by the forcing. An increase in the forcing amplitude results in the narrower spanwise extent of the low-speed region (compare figure \ref{Avfield}$a$ with \ref{Avfield}$b$), as was also reported for the near-wall case \citep{Cassinelli2017}. Finally, it should be mentioned that $\mathbf{\tilde{u}}_s$ is dominated by the streamwise velocity: for example, when $||\mathbf{f}h/u_{\tau,\mathrm{ref}}^2||=1.91$, the maximum streamwise component of $\mathbf{\tilde{u}}_s$ is ${\tilde{u}^+}_{s,max}=4.76$, while its wall-normal counterpart is only ${\tilde{v}^+}_{s,max}=1.01$ (the superscript `$+$' denotes scaling in the inner units). This indicates that the formation of the streaky mean flow is a consequence of the lift-up effect, as explained in detail in \S\ref{ssec:forc}.

Using the computed $\mathbf{\tilde{u}}_s$, two quantities are computed to measure the response to the forcing. The first one is the norm of $\mathbf{\tilde{u}}_s$:
\begin{equation}\label{eq:3.3}
||\mathbf{\tilde{u}}_s||=\sqrt{\frac{2}{V}\int_V \mathbf{\tilde{u}}_s^H\mathbf{\tilde{u}}_s~dV},
\end{equation}
which represents the kinetic energy of the induced motion averaged over the given computational domain. We note that if the forcing amplitude is small and the eddy viscosity in (\ref{eq:2.3}) correctly models the surrounding turbulence, $\mathbf{\tilde{u}}_s$ from (\ref{eq:3.2}) would be identical to $\mathbf{\tilde{u}}$ from the linear model (\ref{eq:2.2}). Although, in practice, this is unlikely to happen due to the crude nature of the eddy-viscosity model, computation of $||\mathbf{\tilde{u}}_s||$ as in (\ref{eq:3.3}) would be useful to assess the linear model (\ref{eq:2.2}), enabling the quantitative comparison of the amplification by the forcing in a real turbulent flow with that in the linear model (\ref{eq:2.2}). The second measure of the response to the forcing is chosen to be the streak amplitude and is defined as
\begin{equation}\label{eq:3.4}
A_s^+=\frac{1}{2}\Big(\max\limits_{y,z}\left[\tilde{u}_s^+(y,z)\right]-\min\limits_{y,z}\left[\tilde{u}_s^+(y,z)\right]\Big),
\end{equation}
where $\tilde{u}_s^+$ is the streamwise component of $\mathbf{\tilde{u}}_s$. This definition follows that used in a laminar boundary layer and previous studies \citep{Andersson2001,Park2011,Alizard2015,Cassinelli2017}.

\begin{figure}
  \centerline{\includegraphics[width=1.1\textwidth]{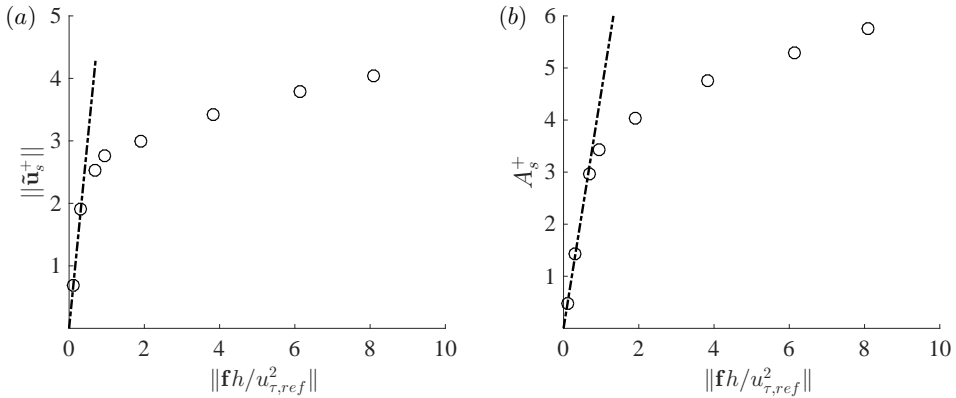}}
  \caption{Flow response to the forcing for S960: (\textit{a}) $||\mathbf{\tilde{u}}_s^+||$; (\textit{b}) $A_s^+$. Here, the definitions of $||\mathbf{\tilde{u}}_s^+||$ and $A_s^+$ are given in (\ref{eq:3.3}) and (\ref{eq:3.4}), respectively.}
\label{LinResp}
\end{figure}

The evolution of the two quantities on increasing the forcing amplitude is shown in figure \ref{LinResp}. As expected, both $\|\mathbf{\tilde{u}}_s^+\|$ and $A_s^+$ increase linearly with $\|\mathbf{{f}}\|$ at small forcing amplitudes (i.e. $\|\mathbf{{f}}h/u_{\tau,\mathrm{ref}}^2\|\lesssim 0.5$). The linear fit is found to follow $\|\tilde{\mathbf{u}}_s^+\|/\| {\mathbf{{f}}}h/u_{\tau,\mathrm{ref}}^2\|\simeq 6.2$. This value is fairly close to $\|\tilde{\mathbf{u}}^+\|/\| {\mathbf{\tilde{f}}}h/u_\tau^2\|=4.0$ obtained with the linear theory based on (\ref{eq:2.2}) (note that $\mathbf{{f}}=\mathbf{\tilde{f}}$), suggesting that the eddy viscosity in (\ref{eq:2.3}) is reasonably a good approximation (see also \S\ref{sec:4.4}). When the forcing amplitude is large enough, neither $\|\mathbf{\tilde{u}}_s^+\|$ nor $A_s^+$ increase linearly anymore, showing a nonlinear behaviour. The largest forcing amplitude considered for the S960 case is $\|{\mathbf{{f}}}h/u_{\tau,\mathrm{ref}}^2\|=8.09$. However, as we shall see in \S\ref{sec:4.1}, this forcing amplitude is too high to study the physical processes in a natural unforced flow. Therefore, forcing amplitude greater than this value is not considered.

\subsection{Streak instability: spectra and rms}\label{ssec:spec}

\begin{figure}
 \centerline{\includegraphics[width=1.1\textwidth]{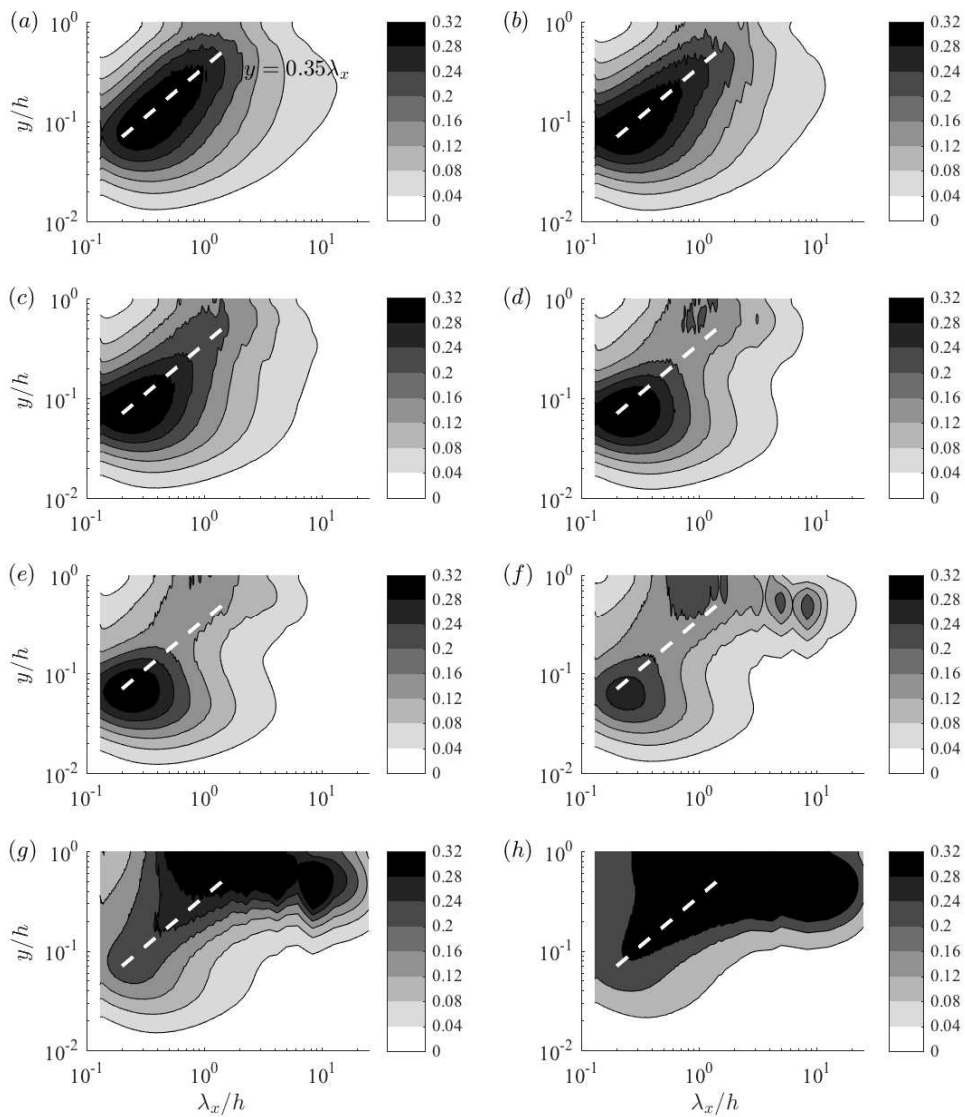}}
 \vspace*{-1.5cm}
  \caption{Premultiplied streamwise wavenumber spectra of wall-normal velocity (S960). (\textit{a}) $A^+_s=0$ (unforced); (\textit{b}) $A^+_s=0.54$; (\textit{c}) $A^+_s=1.44$; (\textit{d}) $A^+_s=2.96$; (\textit{e}) $A^+_s=3.42$; (\textit{f}) $A^+_s=4.04$; (\textit{g}) $A^+_s=4.76$; (\textit{h}) $A^+_s=5.30$. Here, the thick white dashed line indicates $y=0.35 \lambda_x$.}
\label{spectra_comp}
\end{figure}

Now, we investigate the emergence of other flow structures from the driven streamwise-uniform streaks. The focus of this section is given to the detection of wall-normal and spanwise velocity fluctuations at $1<\lambda_x/h <5$, as this is the main statistical feature of LSMs \citep{Hwang2015}. All velocity spectra analyzed in this section have been normalized using the friction velocity of the reference (unforced) simulation. Fig. \ref{spectra_comp} shows the evolution of the streamwise wavenumber spectra of the wall-normal velocity, as the streak (and forcing) amplitude is increased. In the absence of any forcing (figure \ref{spectra_comp}$a$), the spectra are well aligned along the linear ridge $y=0.35 \lambda_x$, as pointed out by \cite{Hwang2015}. At the top end of the linear edge (i.e. $y/h\geq 0.4-0.6$), the spectra show their peak at $\lambda_x/h \simeq 1 - 2$, as in other direct numerical simulations \cite[e.g.][]{Hoyas2006}. Although this streamwise wavelength is a little shorter than $\lambda_x/h \simeq 3 - 4$ observed in the streamwise velocity spectra \cite[][]{Monty2009}, it is evident that this part would correspond to LSMs, given its wall-normal location and streamwise wavelength. Application of the forcing is found to significantly change the wall-normal velocity spectra. Even at fairly small forcing amplitudes (figures \ref{spectra_comp}$b$, $c$), the spectral intensity around the channel centre for $\lambda_x/h \simeq 1 - 2$ is slightly elevated  compared to the unforced case (figure \ref{spectra_comp}$a$). As the forcing amplitude is further increased, the spectra reveal a well-manifested peak at $\lambda_x/h \simeq 1-2h$ and $y/h\geq 0.4-0.6$, consistent with the location of LSMs (figures \ref{spectra_comp}$d$,$e$). At excessively large forcing amplitudes (figures \ref{spectra_comp}$f$-$h$), excitation at larger streamwise length scales is observed. However, the excited streamwise wavelength does not exceed $\lambda_x/h \simeq 10$ at least for $A^+_s<5$ (figure \ref{spectra_comp}$g$).

Through the elevation of the forcing amplitude, it has been consistently observed that the forcing significantly affects the structures at smaller length scales in the near-wall and logarithmic regions. Indeed, the increase in the forcing amplitude gradually reduces the activity in the logarithmic region (figures \ref{spectra_comp}$b$-$e$), significantly weakening even the near-wall processes eventually (figures \ref{spectra_comp}$f$-$h$). It is not clear yet why the application of such a body forcing significantly inhibits the processes at smaller scales in the near-wall and logarithmic regions. However, this feature reminds us of the suppression of the near-wall process, previously observed when a large-scale counter-rotating vortical body forcing is applied at low Reynolds numbers \citep{Schoppa1998,Willis2010,Canton2016}. Indeed, the inhibition of the structures at smaller scales in figure \ref{spectra_comp} is consistent with previous observations. However, in the present study, the suppression of the near-wall and logarithmic regions did not yield any skin-friction reduction, unlike previous studies at low Reynolds numbers. This is because the forcing itself has elevated the skin-friction generation process at the scale of the applied forcing via the enhanced lift-up effect (see also the discussion in \S\ref{sec:3.1}). Finally, we note that the observed suppression of the processes at smaller scales is clearly due to an interaction of the forcing with other scales. Therefore, further discussion on this issue is beyond the scope of the present study.

\begin{figure}
  \centerline{\includegraphics[width=1.1\textwidth]{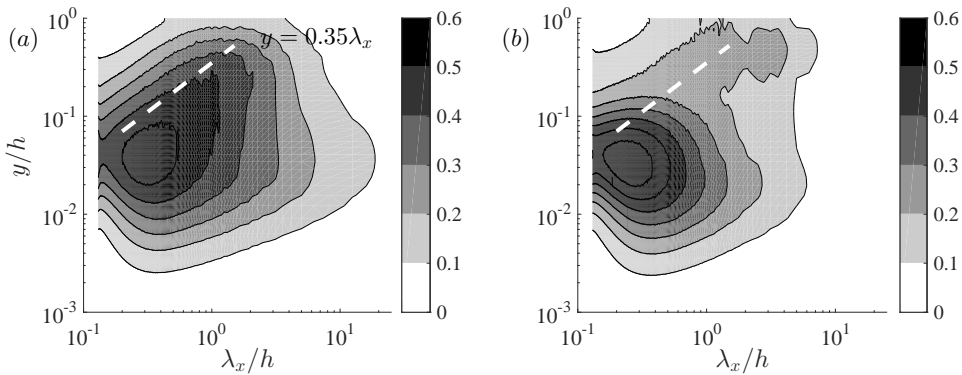}}
  \caption{Premultiplied streamwise wavenumber spectra of spanwise velocity (S960): (\textit{a}) $A^+_s=0$ (unforced); (\textit{b}) $A_s^+=3.42$. Here, the thick white dashed line indicates $y=0.35 \lambda_x$.}
\label{spectra_base}
\end{figure}

The behaviour of the spanwise velocity spectra on increasing the forcing amplitude has also been found to be qualitatively the same as that of the wall-normal velocity spectra. Fig. \ref{spectra_base} shows the streamwise wavenumber spectra of the spanwise velocity for the unforced and forced cases. Similar to the wall-normal velocity spectra, the forcing resulting in $A_s^+\simeq3-4$ excites the spectral intensity at $\lambda_x/h\simeq1-2$ and $y/h\geq0.4-0.6$, while weakening the activity of the structures in the near-wall and logarithmic regions.

The observation made with figures \ref{spectra_comp} and \ref{spectra_base} suggests that an intense cross-streamwise velocity structure at $\lambda_x/h \simeq1-2$ is excited at low amplitudes of the body forcing. It is important to remember that the applied forcing is uniform in the streamwise direction ($\lambda_x=\infty$), thus this cross-streamwise velocity structure at a finite streamwise length scale should not be the direct outcome of the forcing. Instead, it suggests that another physical mechanism is at play in the generation of this structure. In this respect, it is encouraging to compare the observed streamwise wavelength of the cross-streamwise velocity structure with that of the streak instability in the previous theoretical analysis. The stability analysis using an eddy-viscosity model \cite[]{Park2011} predicted that a streaky base flow with $\lambda_z/h=4$ becomes unstable at $\lambda_x/h\simeq3-6$. At first glance, this prediction does not seem to match well with $\lambda_x/h\simeq1-2$ in the present numerical experiment. However, the spanwise spacing of the streak in the present study is $\lambda_x/h\simeq1.5$, considerably smaller than $\lambda_z/h=4$ in \cite{Park2011}. Indeed, a recent extension of the theoretical analysis to smaller spanwise wavelengths showed that there exists a self-similar relation between the spanwise wavelength of the imposed streak and the streamwise wavelength of its instability \citep{Alizard2015}. If this is adopted, a streak at $\lambda_z/h \simeq 1.5$ would become unstable at $\lambda_x/h \simeq 1.13-2.25$, showing a fairly good agreement with $\lambda_x/h \simeq 1-2$ observed in the present study.

\begin{figure}
  \centerline{\includegraphics[width=1.05\textwidth]{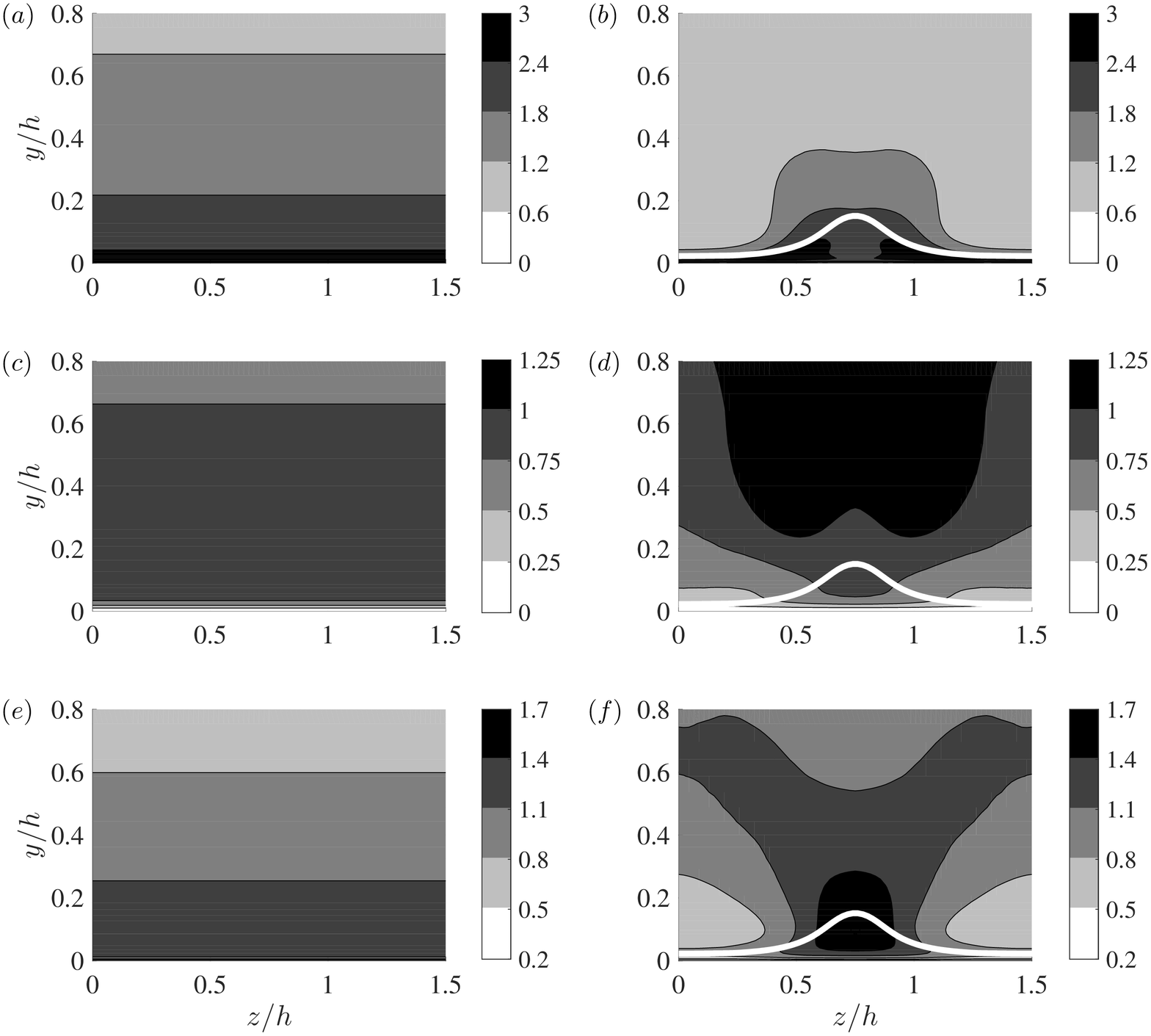}}\vspace*{-5mm}
  \caption{Cross-streamwise view of the rms of the velocities} for $(a,c,e)$ $A_s^+=0$ and $(b,d,f)$ $A_s^+=3.42$ (S960): (\textit{a},\textit{b}) streamwise velocity; (\textit{c},\textit{d}) wall-normal velocity; (\textit{e},\textit{f}) spanwise velocity. The thick white line indicates the location where the cross-streamwise mean velocity is $0.9U_c$ ($U_c$ is the centreline velocity of the simulation).
\label{fluc}
\end{figure}

Further evidence of the streak instability mechanism is found in the cross-streamwise view of the velocity fluctuations in figure \ref{fluc}, where the left and right columns visualise the unforced and forced cases, respectively. We remind the reader that the streaky mean flow shows an even symmetry about $z/h=0.75$ (figure \ref{Avfield}) and that the eigenstructure of a sinuous-mode streak instability is mathematically defined by odd symmetry of the streamwise and wall-normal velocities and by even symmetry of the spanwise velocity about this axis \cite[see e.g. ][]{Andersson2001}: i.e.
\begin{eqnarray}
\check{u}(y,z^*)&=&-\check{u}(y,-z^*), \nonumber\\
\check{v}(y,z^*)&=&-\check{v}(y,-z^*), \nonumber \\
\check{w}(y,z^*)&=&\check{w}(y,-z^*),
\end{eqnarray}
with $z^*=z-0.75h$. Here, $\check{\mathbf{u}}(y,z^*)=(\check{u},\check{v},\check{w})$ is the eigenfunction of the streak instability, and the instability mode in physical space is constructed as $\check{\mathbf{u}}(y,z^*)e^{ik_{x,m} x}$ ($k_{x,m}$ is the streamwise wavenumber of the instability). Given the mathematical definition, the streamwise and wall-normal velocity fluctuations solely from the sinuous-mode instability should be zero along $z/h=0.75$ and the spanwise velocity fluctuation is supposed to have a local maximum/minimum along this axis. Therefore, in the presence of background turbulence, the streamwise and wall-normal velocity fluctuations of the sinuous-mode instability would always show a local minimum along $z/h=0.75$, while the spanwise one should exhibit a local maximum/minimum along this axis. These features indeed appear in the computed cross-streamwise view of turbulent velocity fluctuations in figure \ref{fluc}: both the streamwise and wall-normal velocity fluctuations exhibit two symmetric peaks with a local minimum given along $z=0.75h$ (figures \ref{fluc}$b$,$d$), while the spanwise one shows a local maximum along $z/h=0.75$ (figure \ref{fluc}$f$). We also note that the cross-streamwise view of the velocity fluctuations is qualitatively the same as the one in a laminar boundary layer \citep{Andersson2001}, indicating the presence of a sinuous-mode instability. In \S\ref{ssec:DMDre}, we will provide further evidence of the sinuous-mode instability by visualising the most energetic DMD mode (see figure \ref{InstvdDMD}).

\subsection{Dynamic mode decomposition}{\label{ssec:DMDre}}

A DMD is performed to examine the flow structure in detail. For the snapshots of the DMD, a time series of a specific Fourier component of the velocity field in the lower half of the wall-normal domain is taken from the simulation with a sufficiently large streak amplitude. The spanwise wavenumber is chosen to be the same as that of the forcing ($k_{z}=2\pi/\lambda_z$ with $\lambda_z/h=1.5$), while a range of different streamwise wavenumbers ($k_x=2\pi/\lambda_x$) are considered ($2<\lambda_x/h<5$). For given $k_z$ and $k_x$, each snapshot is constructed such that:
\begin{equation}\label{eq:3.5}
\bar{\mathbf{u}}(y,z;k_x)=\hat{\mathbf{u}}(y;k_x,k_z)e^{ik_zz}+\hat{\mathbf{u}}(y;k_x,-k_z)e^{-ik_zz}.
\end{equation}
We note that $\bar{\mathbf{u}}(y,z;k_x)$ is the streamwise Fourier component given only with $\pm k_z$ spanwise Fourier components. The DMD modes and the corresponding eigenvalues $\mu_j$ are then computed following the procedure in \S\ref{ssec:DMD}. The time interval, the number of POD modes and the number of snapshots for the DMD are carefully chosen to ensure good temporal resolution of large-scale dynamics: $\Delta t u_\tau/h=0.009$, $r=30$ and $N= 1000$ (see Appendix \ref{DMDapp} for a detailed discussion on the choice of parameters). Once the DMD is performed, the associated angular frequency is computed by taking the real part of $\omega_j=\mathrm{i}\ln \mu_j/\Delta t$. Since the snapshot in (\ref{eq:3.5}) represents a streamwise Fourier component, the downstream propagating speed of the DMD mode is computed by
\begin{equation}\label{eq:3.6}
c_j=\frac{\omega_j}{k_x},
\end{equation}
where $c_j$ is the phase speed of the $j$-th DMD mode. The relative amplitude $\alpha_j$ of all of the DMD modes is finally determined (see also \S\ref{ssec:DMD}).

\begin{figure}
 \centerline{\includegraphics[width=0.85\textwidth]{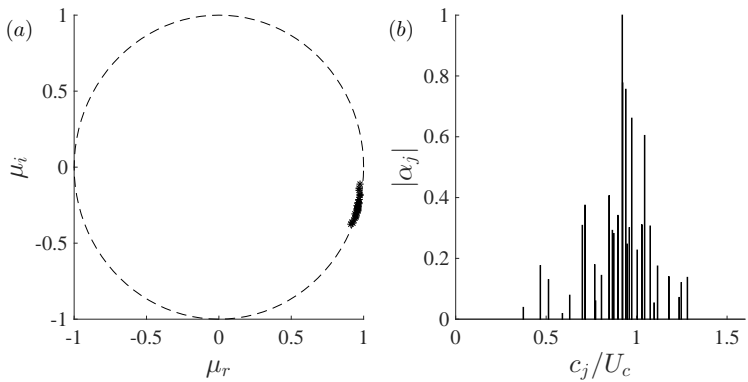}}
  \caption{DMD analysis with $\lambda_x=3.57h$ for $A_s^+=3.42$ (S960): (\textit{a}) eigenspectra in the $\mu_r$-$\mu_i$ plane; (\textit{b}) mode amplitude $|\alpha_j|$ with phase speed $c_j$ (the mode amplitude is normalised by the largest one).}
\label{fig:unitc}
\end{figure}

\begin{figure}
  \centerline{\includegraphics[width=0.80\textwidth]{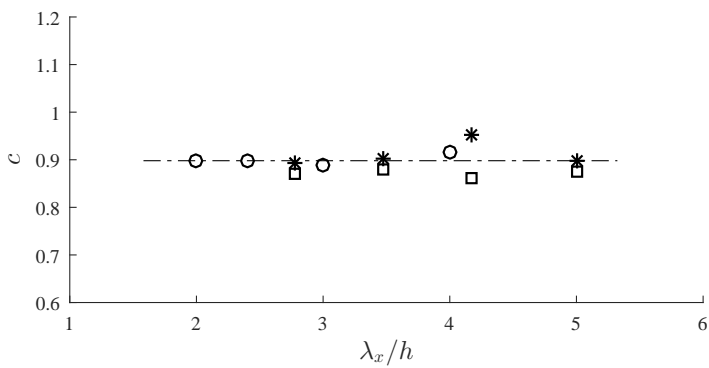}}
  \caption{Variation of phase speed ($c/U_c$) with streamwise wavelength: {\tiny$\square$}, $A^+_s=4.11$ (S490), $*$, $A^+_s=4.04$ (S960); $\circ$, $A^+_s=3.77$ (S2050). Here, the horizontal line indicates $c=0.9U_c$.}
\label{ph_speed}
\end{figure}

The DMD result of the S960 simulation is shown in figure \ref{fig:unitc} for $\lambda_x/h=3.57$ and $A_s^+=3.42$. The DMD modes are deemed to be neutral, given that the eigenvalues $\mu_j$ sit well on the unit circle (figure \ref{fig:unitc}$a$). This also indicates that the snapshots are indeed collected from a statistically stationary flow with a good resolution. Their asymmetry is due to the processing of complex-valued snapshots instead of real-valued ones. Using (\ref{eq:3.6}), the computed eigenvalues are subsequently transformed to the phase speed of the DMD modes. The contribution of each DMD mode is then assessed by computing its amplitude $\alpha_j$, as shown in figure \ref{fig:unitc}$(b)$. The phase speed of the DMD modes appears mainly in the range of $0.7\lesssim c/U_c\lesssim 1.1$ ($U_c$ is the centreline velocity of each simulation), and, in particular, the most energetic DMD mode exhibits $c/U_c\simeq 0.9$.

It is important to mention that the computed phase speed of the most energetic DMD mode is found to not significantly depend on the streamwise wavelength and the Reynolds number, as long as the streak amplitude is chosen to be $A^+_s>3$. Fig. \ref{ph_speed} shows the phase speed of the most energetic DMD mode with respect to the streamwise wavelength for all the considered Reynolds numbers. It appears that the most energetic DMD mode, for different streamwise lengths and Reynolds numbers, robustly exhibits $c/U_c\simeq 0.9$. This suggests that the propagation speed of the flow structure generated by the streak instability scales well with the centreline velocity, consistent with \cite{delAlamo2009} who showed the outer scaling of the propagation speed of large-scale structures.

\begin{figure}
 \centerline{\includegraphics[width=1.15\textwidth]{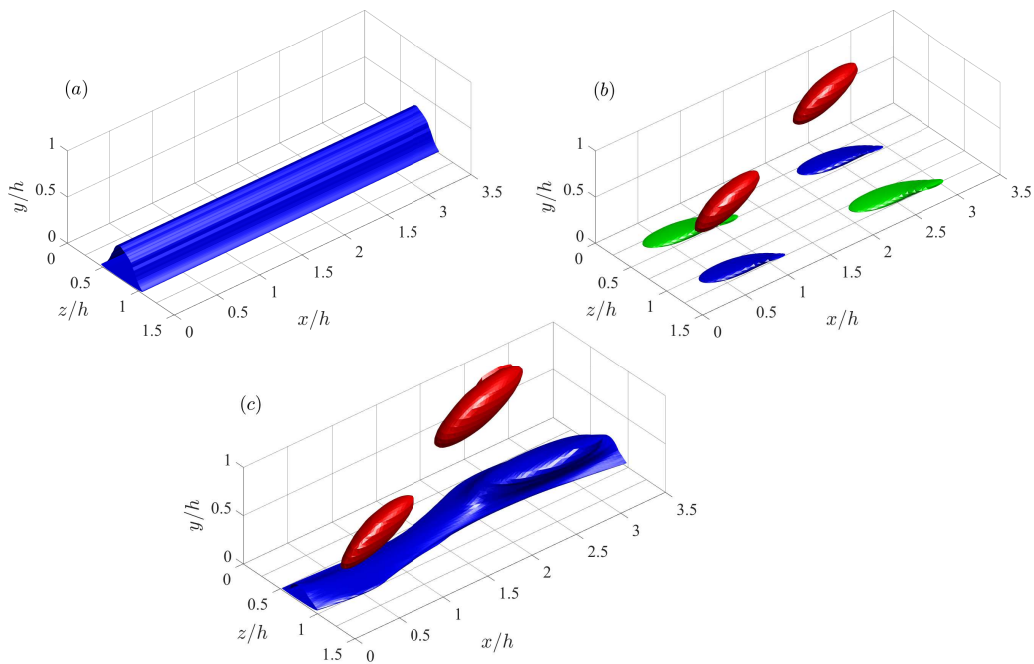}}
  \caption{Visualisation of the streak instability using the most energetic DMD mode from the S960 simulation ($\lambda_x/h=3.73$ and $A_s^+=4.73$): $(a)$ the low-speed streak with $u_s'=-0.03U_c$ ($u_s'\equiv u_s(y,z)-\langle{u}_s(y,z)\rangle_{z}$ where $\langle \cdot \rangle_z$ denotes average in the spanwise direction); $(b)$ the DMD mode $\mathbf{u}_{dmd}$ where the green iso-surfaces are a positive streamwise velocity $(0.8\max[u_{dmd}])$, the blue ones a negative streamwise velocity $(u=0.8\min[u_{dmd}])$, and the red ones a negative wall-normal velocity $(v=0.8\min[v_{dmd}])$; $(c)$ the low-speed streak with the DMD mode ($\mathbf{u}_s-\langle\mathbf{u}_s\rangle_{z}+\gamma\mathbf{u}_{dmd}$ where $\gamma$ is an appropriate tuning constant for visualisation). In $(c)$, the blue iso-surface indicates a negative streamwise velocity, while the red ones indicate a negative wall-normal velocity.}
\label{InstvdDMD}
\end{figure}

Finally, the most energetic DMD mode is examined. Fig. \ref{InstvdDMD} ($a$) and ($b$) show the streaky uniform mean flow and the most energetic DMD mode for $\lambda_x/h=3.73$, respectively, and their combination is given in figure \ref{InstvdDMD} ($c$). Fig. \ref{InstvdDMD} ($c$) clearly reveals that the structure is composed of a streamwise meandering low-speed streak (blue isosurface in figure \ref{InstvdDMD}$c$) and a streamwise-alternating pattern of wall-normal velocity structures (red iso-surface in figure \ref{InstvdDMD}$c$). We note that this is a robust feature of the computed DMD mode and it does not significantly depend upon the choice of the DMD parameters, such as the streamwise wavelength $\lambda_x$, the number of snapshots $N$, and the number of POD modes $r$. Therefore, the spatial structure of the DMD mode in figure \ref{InstvdDMD} clearly indicates that the flow structure with an intense cross-streamwise turbulent kinetic energy in \S\ref{ssec:spec} is directly linked to the sinuous-mode streak instability. In this respect, it is finally worth mentioning that the structure of the most energetic DMD mode with the streaky mean flow highly resembles the general form of the traveling-wave solutions in \cite{Hwang2016b}, which is a mathematical representation of the self-sustaining process.

\section{Discussion}\label{sec:Dis}
Thus far, we have explored the instability mechanism of an `amplified' streak in the outer region (VLSM) and its mode as the initiating physical process of LSMs. A body forcing, designed to drive an infinitely long streak \citep{Hwang2010b}, is implemented in a set of numerical simulations. As the forcing amplitude is increased, a long streaky structure, which resembles a VLSM, emerges. It is shown that the presence of such a long streaky structure in the outer region generates an intense cross-streamwise velocity structure at $\lambda_x/h\simeq 1-2$, and this streamwise length scale shows good agreement with that predicted by the previous stability analysis \cite[]{Park2011,Alizard2015}. The cross-streamwise view of the turbulence statistics has also revealed that this structure has features mathematically consistent with the sinuous-mode streak instability. Finally, a DMD analysis shows that this structure is characterised by a streamwise meandering motion of the driven streak and alternating cross-streamwise velocity structures, indicating that the cross-streamwise velocity structure at $\lambda_x/h\simeq 1-2$ is a consequence of the sinuous-mode streak instability.

\begin{figure}
  \centerline{\includegraphics[width=0.65\textwidth]{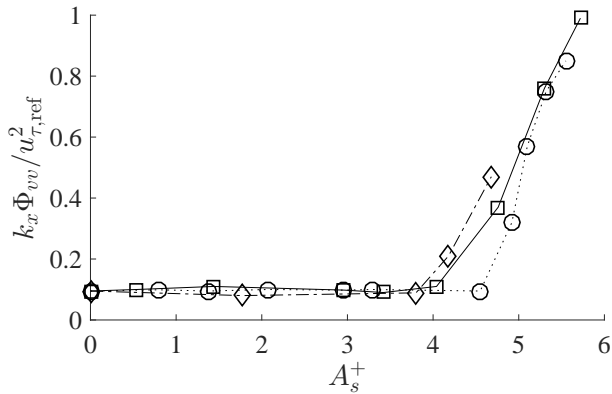}}
  \caption{Premultiplied streamwise wavenumber spectral intensity of wall-normal velocity with the streak amplitude at $y/h=0.64$ for $\lambda_x/h=3.57$: $\circ$, S490; {\tiny$\square$}, S960; $\diamond$, S2050.}
\label{verticalspectra}
\end{figure}

The present numerical experiment is specifically designed to observe the streak instability, while taking the nonlinearity of the governing equation and background turbulence fully into account. This is achieved by artificially driving a streak with the optimal forcing, but we note that this does not allow the driven streak to break down. For this reason, in the present numerical experiment, the nonlinearly evolved streak instability and the accompanying vortical structures coexist. Therefore, with the present numerical experiment alone, it is difficult to precisely explain the causal relation between the streak instability and the vortical structures. However, as we shall see in \S\ref{sec:4.3}, the sinuous-mode streak instability essentially originates from the spanwise shear created by the presence of the driven large-amplitude streak. This implies that the streak instability is the seeding mechanism of the vortical structures, and this interpretation is consistent with \cite{Hwang2016}, where the emergence of the streak instability prior to the vortical structures is clearly identified with the minimal-unit simulation.

In many ways, the flow structure emerging at $\lambda_x/h\simeq 1-2$ via the streak instability is remarkably similar to the LSM. Firstly, its streamwise length scale and wall-normal location (figures \ref{spectra_comp} and \ref{spectra_base}) are in good agreement with those of the LSM \citep{Kovasznay1970,Monty2009}. Secondly, the structure is mainly featured by intense wall-normal and spanwise turbulent kinetic energy, as the LSM is in a real flow \citep{Hwang2015}. Thirdly, a series of the cross-streamwise velocity components of this structure are collectively aligned along the low-speed streak (figure \ref{InstvdDMD}$c$), consistent with the early view of a VLSM as a concatenation of adjacent LSMs \cite[e.g.][]{Kim1999,Baltzer2013}. However, it should be mentioned that the present numerical experiment provides a different interpretation of the relation between the LSM and the VLSM: the reason that a VLSM appears as a concatenation of a number of LSMs is because a series of the LSMs are formed along a long VLSM by the streak instability. Lastly, the emergence of the sinuous-mode instability implies that the VLSM would meander along the streamwise direction (figures \ref{Inststreak} and \ref{InstvdDMD}), as previously observed in \cite{Hutchins2007}.

All of these observations suggest that the LSM is initiated by the sinuous-mode instability of the amplified streak in the outer region (VLSM). This interpretation is well integrated into the self-sustaining nature of the large-scale structures composed of VLSMs and LSMs \citep{Hwang2010d,Hwang2015,Hwang2016,Hwang2016b}, and is also consistent with the previous theoretical studies on the streak instability \citep{Park2011,Alizard2015}. There have been several different propositions on the formation mechanism of the LSM, such as the mergers/growth mechanism \citep{Perry1986,Zhou01,Adrian2007} and the Orr mechanism \citep{Jimenez2013a,Jimenez2013}. However, it should be stressed that, to the best of our knowledge, only the streak instability mechanism currently explains the streamwise length-scale determination of the LSM among these different propositions.

\subsection{Emergence of streak instability at large scales}\label{sec:4.1}
In the near-wall region, it has been shown that the streak instability is subcritical, admitting strong transient growth at low streak amplitudes before the onset of a normal-mode instability \cite[]{Schoppa2002}. Consistent with this, our recent work has shown that the streak instability and the related near-wall quasi-streamwise vortices at $\lambda_x^+\simeq 200-300$ spontaneously emerge on increasing the streak amplitude \citep{Cassinelli2017}, exhibiting a sensitive response to the initial condition and/or external noise, as is typically observed in a highly non-normal system \citep{Schmid2001}. However, it should also be mentioned that such non-normal growth of the perturbation around a low-amplitude streak is typically an outcome of the interaction with the marginally stable sinuous mode. Therefore, in practice, distinguishing the transient growth from the normal-mode instability is almost impossible at least for near-wall turbulence and for transitional laminar boundary layers, as pointed out by \cite{Hepffner2005}.

\begin{figure}
  \centerline{\includegraphics[width=1.1\textwidth]{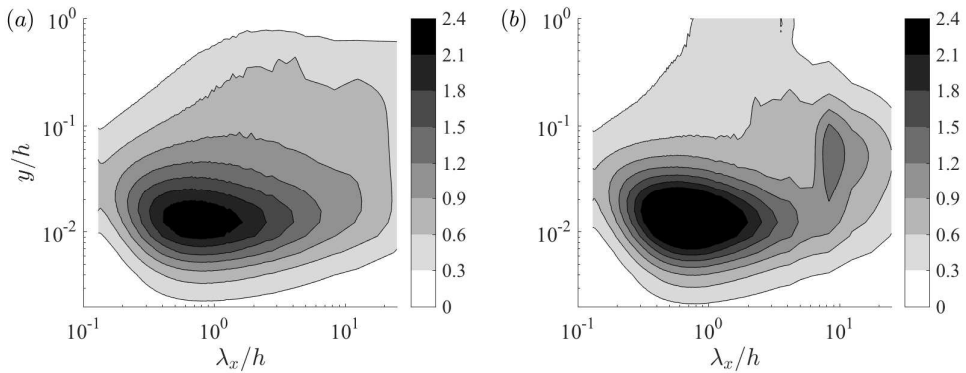}}
  \caption{Premultiplied streamwise wavenumber spectra of streamwise velocity (S960); (\textit{a}) unforced case; (\textit{b}) $A_s^+=4.04$.}
\label{spectra_base_stream}
\end{figure}

To carefully examine the emergence of the streak instability in the present numerical experiment, the one-dimensional spectral intensity of the wall-normal velocity for $\lambda_x/h=3.57$ is plotted in figure \ref{verticalspectra} against the streak amplitude. For relatively low streak amplitudes ($A_s^+<4$), the intensity of the wall-normal velocity is very slightly elevated. However, when the streak amplitude reaches $A_s^+\simeq 4$, the spectral intensity drastically increases for all the Reynolds numbers considered. This behaviour is very interesting, as it indicates that the onset of a normal-mode instability would be $A_s^+\simeq 4$. However, this does not imply that the streaky motion in a real flow (VLSM) would experience such a normal-mode instability for $A_s^+ \gtrsim 4$. Firstly, as is evident in figure \ref{verticalspectra}, the typical spectral intensity of the wall-normal velocity in the unforced simulation is much smaller than that observed for $A_s^+ \gtrsim 4$. Secondly, the spectra of the wall-normal velocity in the unforced simulation do not exhibit high intensity for $\lambda_x>5-10 h$ in the outer region (figure \ref{spectra_comp}$a$). However, the forced simulations with $A_s^+ \gtrsim 4$ show much higher spectral intensity in this range of the streamwise wavelength (figures \ref{spectra_comp}$f$-$h$). Finally, the forced simulations with relatively low streak amplitudes ($A_s^+ \lesssim 4$) are not actually unresponsive to the surrounding perturbations: the wall-normal velocity spectra of the simulations with $3 \lesssim A_s^+ \lesssim 4$ also exhibit a fairly energetic peak at $\lambda_x/h \simeq 1-2$ (figures \ref{spectra_comp}$d$-$f$), which would be more consistent with the spectra of the unforced reference simulation.

These observations suggest that VLSMs in a real flow are unlikely to reach $A_s^+\gg4$ because they should break down quickly with the emergence of the streak instability \cite[]{Hwang2016}. The streak amplitude would reach at best $A_s^+\simeq 4$, below which the wall-normal velocity structure is excited primarily at $\lambda_x/h \simeq 1-2$. This also indicates that the streak instability process in the outer region may be mainly driven by a non-modal mechanism (secondary transient growth), as that in the near-wall region. Finally, it should be stressed that the wall-normal velocity structure at $\lambda_x/h \simeq 1-2$ is the first feature noticeable statistically on increasing the streak amplitude, and it depends very little on the type of body forcing that drives the streak. This issue will be further discussed in \S\ref{sec:4.4}.

\begin{figure}
  \centerline{\includegraphics[width=1.15\textwidth]{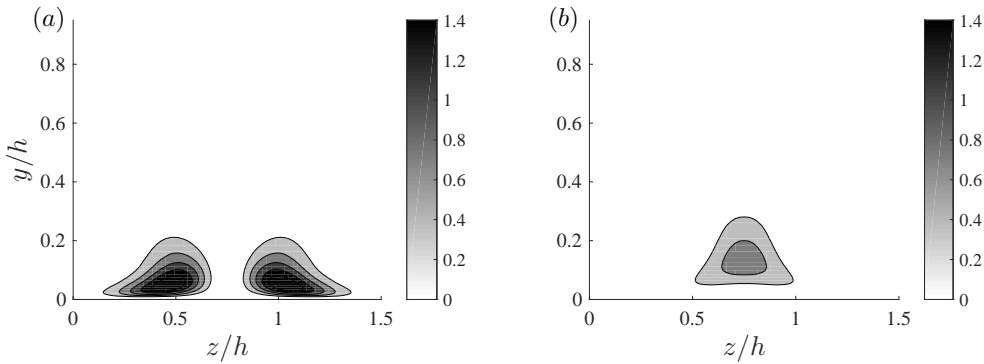}}
  \caption{Turbulent kinetic-energy production around the streaky mean flow for $A^+_s=3.42$ (S960): (\textit{a}) $-\langle u'w'\rangle_{x,t}\partial \overline{U}^+(y,z)/\partial z$; (\textit{b}) $-\langle u'v'\rangle_{x,t}\partial\overline{U}^+(y,z)/\partial y$.}
\label{shear}
\end{figure}

\subsection{The streamwise length scale of very large-scale motion}\label{sec:4.2}
The present numerical experiment also provides important physical insight into the streamwise length scale determination of the VLSM. If the streak amplitude is given by $3 \lesssim A_s^+ \lesssim 4$ just before the breakdown phase in the self-sustaining process (see also the discussion in \S\ref{sec:4.1}), the largest streamwise wavelength of the wall-normal velocity structure generated by the streak instability mechanism would be limited to $\lambda_x/h \simeq 10-20$ (see figures \ref{spectra_comp}$d$-$f$). The generated wall-normal velocity structure would then trigger linear amplification of the streaky structures that mainly carry the streamwise turbulent kinetic energy (i.e. lift-up effect). As discussed in \S\ref{ssec:forc}, the largest possible linear amplification of the streaks takes place at infinitely long streamwise wavelength and it generally prefers longer streamwise length for larger amplification  \cite[]{delAlamo2006a,Pujals2009,Cossu2009,Hwang2010a,Willis2010}. Therefore, the largest admissible streamwise wavelength of the streak instability (i.e. $\lambda_x/h \simeq 10-20$) is expected to determine the streamwise length scale of the VLSM.

This scenario is examined in figure \ref{spectra_base_stream}, where the streamwise wavenumber spectra of the streamwise velocity for both the unforced and forced simulations are shown. Here, the streak amplitude of the forced simulation is chosen to be  $A_s^+=4.04$, as the excitation of the wall-normal velocity structure around $\lambda_x/h \simeq 10$ is well visible at this amplitude (see figure \ref{spectra_comp}$f$). As expected from the wall-normal and spanwise velocity spectra (figures \ref{spectra_comp} and \ref{spectra_base})), the streamwise velocity spectra of the forced simulation reveal elevated spectral intensity at $\lambda_x/h \simeq 1-2$ and $y/h \gtrsim 0.3$ (figure \ref{spectra_base_stream}$b$), the location associated with the LSM (figure \ref{spectra_base_stream}$a$). However, in the streamwise velocity spectra of the forced simulation (figure \ref{spectra_base_stream}$b$), there exists another more distinct peak at $\lambda_x/h \simeq 10$ throughout the logarithmic layer. Importantly, the wall-normal location and the streamwise wavelength of this peak are very close to those of the VLSM, directly supporting the notion that the largest streamwise wavelength of the streak instability determines the streamwise length scale of the VLSM.

\subsection{Physical mechanism of streak instability}\label{sec:4.3}

\begin{figure}
  \centerline{\includegraphics[width=1.05\textwidth]{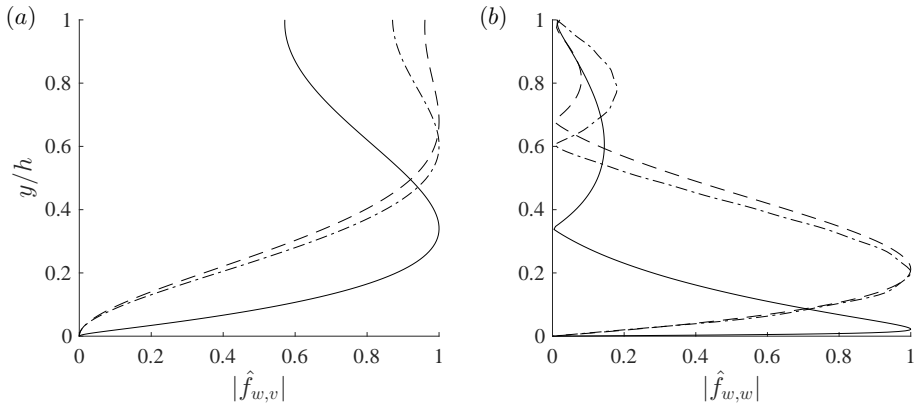}}
  \caption{The wall-normal profile of the optimal forcing with three different eddy viscosities in (\ref{eq:2.2}): (\textit{a}) $|\hat{f}_{w,v}|$, (\textit{b}) $|\hat{f}_{w,w}|$. Here, \protect \solid, the Cess eddy viscosity ($\nu_t=\nu_t(y)$ in (\ref{eq:2.3})); \protect \dashdot, a constant eddy viscosity ($\nu_t=\max_y \nu_t(y)$); \protect \dashed, no eddy viscosity ($\nu_t=0$).}
\label{fprofile}
\end{figure}

The second-order cross-streamwise statistics and the DMD analysis suggest that the vortical structure at $\lambda_x/h\simeq 1-2$ originates from the sinuous-mode streak instability in the form of a streamwise meandering streak and alternating cross-streamwise velocity components (figure \ref{InstvdDMD}$c$). The sinuous-mode instability is understood to be an energy-production process from the spanwise mean shear via the inflectional mechanism \cite[e.g.][]{Park1995}. To investigate this feature of the sinuous-mode instability in the present numerical experiment, turbulent production by the streaky mean flow are further examined in this section. The streaky mean flow is defined as $\overline{U}(y,z)=U_0(y)+\tilde{u}(y,z)$ because the mean flow is dominated by the streamwise component (see the discussion in \S\ref{sec:3.1}). Turbulent production around such a mean flow are given by $-\langle u'w'\rangle_{x,t}\partial \overline{U}^+/\partial z$ and $-\langle u'v'\rangle_{x,t} \partial\overline{U}^+/\partial y$; the former indicates production by the spanwise mean shear, while the latter is production by the wall-normal mean shear. The two turbulent kinetic-energy production mechanisms are visualised in figure \ref{shear}. Production by the spanwise shear appears at either side of the low-speed region at $z/h=0.75$ (figure \ref{shear}$a$), whereas that by the wall-normal shear is strong only in the top of the low-speed region (figure \ref{shear}$b$). In this figure, it is evident that production by the spanwise shear is much grater than that by wall-normal shear, confirming that the streak instability in the present study is a sinuous mode.

\subsection{Robustness to the forcing profile}\label{sec:4.4}

\begin{figure}
 \centerline{\includegraphics[width=1.1\textwidth]{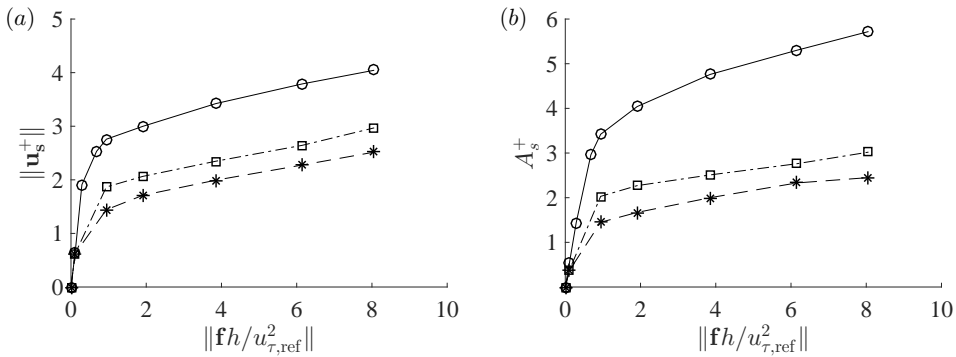}}
  \caption{Flow response to the forcing for S960: (\textit{a}) $||\mathbf{\tilde{u}}_s^+||$; (\textit{b}) $A_s^+$. Here, {\tiny$\bigcirc$}, the Cess eddy viscosity ($\nu_t=\nu_t(y)$ in (\ref{eq:2.3}));  $\ast$, no eddy viscosity ($\nu_t=0$); {\tiny$\square$}, a constant eddy viscosity ($\nu_t=\max_y \nu_t(y)$).}
\label{fig:forcnorm}
\end{figure}

Thus far, the optimal forcing employed in the present study is obtained using the linearised Navier-Stokes equation with the eddy viscosity in (\ref{eq:2.3}). However, it should be pointed out that this forcing is not the only type of the forcing that can generate a streak. Indeed, the only requirement for body forcing to generate a streak is to contain reasonably strong cross-streamwise components, given the mechanism of the streak formation (i.e. the lift-up effect). In this section, we therefore consider two additional forcing profiles to examine the robustness of the present result: one is the optimal forcing computed without eddy viscosity ($\mathbf{\nu}_t=0$), and the other is obtained by setting $\nu_t=\max_y(\nu_t(y))$. We note that these different eddy viscosities are introduced only for the linear theory in order to generate different types of forcing. Unsurprisingly, the optimal forcing with the two different eddy viscosities are also found to consist of a pair of counter-rotating streamwise vortices (not shown; see also the discussion in \S\ref{ssec:forc}). However, as shown in figure \ref{fprofile}, their wall-normal profiles significantly differ from that of the original optimal forcing. While the wall-normal profiles of the two new forcings are not very different from each other, they have much larger values in the location much further from the wall: indeed, the wall-normal component of the two forcings show a maximum at $y/h\simeq 0.6$, whereas the optimal forcing with the original Cess viscosity exhibits its peak at a location much closer to the wall, $y/h\simeq0.35$.

\begin{figure}
  \centerline{\includegraphics[width=1.1\textwidth]{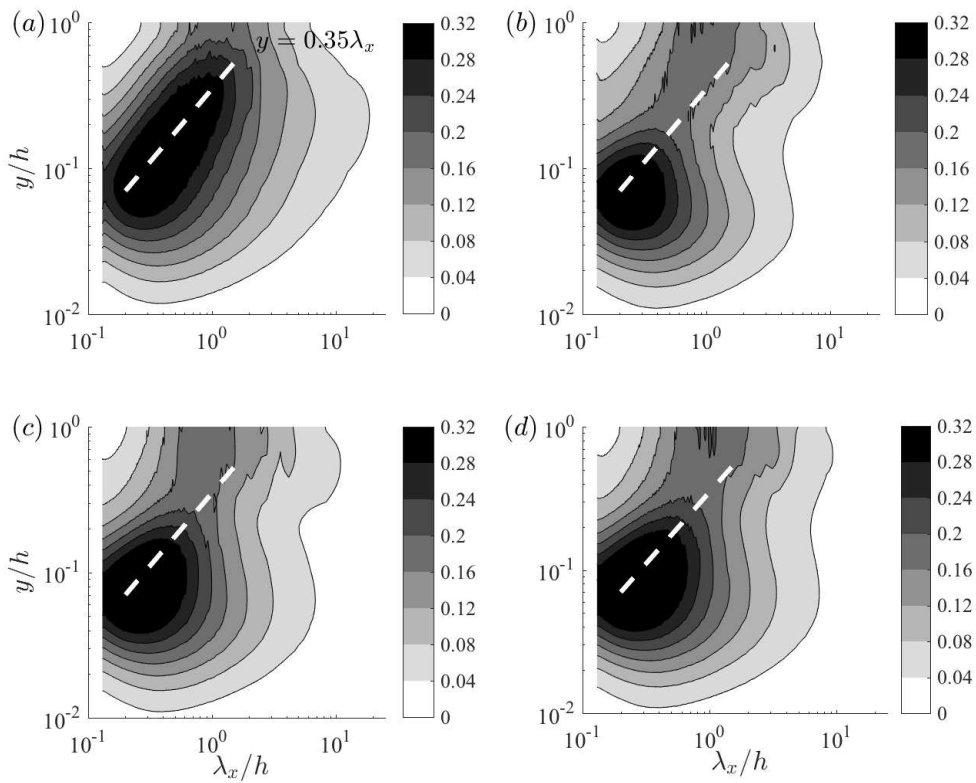}}
  \caption{Premultiplied streamwise wavenumber spectra of wall-normal velocity (S960). (\textit{a}) $A^+_s=0$ (unforced); (\textit{b}) $A^+_s=3.42$ with the Cess eddy viscosity ($\nu_t=\nu_t(y)$ in (\ref{eq:2.3})); (\textit{c}) $A^+_s=1.47$ with no eddy viscosity ($\nu_t=0$); (\textit{d}) $A^+_s=2.04$ with a constant eddy viscosity ($\nu_t=\max_y \nu_t(y)$). Here, the thick white dashed line indicates $y=0.35 \lambda_x$.}
\label{viscspectra}
\end{figure}

The response to the three different types of the forcing was then examined, as reported in figure \ref{fig:forcnorm}. While all the three forcings are found to robustly generate the streaky motion (not shown; see also the discussion in \S\ref{ssec:forc}), the optimal forcing obtained with the Cess eddy viscosity gives rise to a significantly larger response of the streamwise velocity component. This implies that the linearised Navier-Stokes equation with the Cess eddy viscosity provides a much better description of the linear process in a real turbulent flow than the other two cases with different eddy viscosities. This also suggests that the streaky motion generated by the optimal forcing with the Cess eddy viscosity would be the closest to the one in a real flow, because it implies that such a streaky motion can be most easily generated by the nonlinear term.

Here, it should be stressed that this interesting behaviour originates from an important physical feature of the Cess eddy viscosity, which satisfies the following mixing length hypothesis:
\begin{equation}\label{eq:4.1}
\nu_t\frac{dU}{dy}=-\overline{u'v'}.
\end{equation}
In the logarithmic region, $dU/dy \sim 1/y$ and the Reynolds shear stress is approximately constant. Therefore, the considered $\nu_t(y)$ should grow linearly with $y$ in the logarithmic region according to (\ref{eq:4.1}). It should be pointed out that this tendency of $\nu_t$ must be generic in any turbulence model for wall-bounded shear flows. In the near-wall region, both the integral and dissipation length scales are the inner length scale $\nu/u_\tau$, indicating that the dissipation process should be dominated by the molecular viscosity. On the other hand, in the outer region, the integral length scale ($h$) is largely separated from the dissipation length scale ($(\nu^3 h/u_{\tau}^3)^{1/4}$), and this length-scale separation is given by the ratio of the integral scale to the dissipation scale, i.e. $Re_\tau^{3/4}$. Therefore, in the outer region, the energy dissipation takes place through the Richardson-Kolmogorov cascade, requiring large $\nu_t$ from the modelling viewpoint. In the logarithmic region, the only way to smoothly connect the two very different dissipation mechanisms in the near-wall and outer regions is by having monotonically increasing $\nu_t$ in the wall-normal direction, as reflected in (\ref{eq:4.1}). This behaviour of $\nu_t$ explains why the optimal forcing obtained with the Cess eddy viscosity has large values in the region close to the wall -- the large value of the forcing should generate a large response due to the small eddy viscosity in this region. It has been recently shown that this behaviour of $\nu_t$ is essential for understanding the scaling of coherent structures in the logarithmic and outer regions \citep{Hwang2016a}: for example, the Reynolds-number-dependent scaling of the peak wall-normal location of streamwise turbulence intensity, the near-wall penetration of the large-scale outer structures, and the incomplete self-similarity of the streamwise and spanwise velocity components of the coherent structures in the logarithmic region.

By applying the three different types of optimal forcing in figure \ref{fig:forcnorm} to a numerical simulation, the sensitivity of the streak instability was finally examined. Fig. \ref{viscspectra} shows the streamwise wavenumber spectra of the wall-normal velocity when  each of the three optimal forcings just begins to exhibit streak instability. Here, the forcing amplitude follows the definition of (\ref{eq:3.1}), and the corresponding responses are evaluated using (\ref{eq:3.3}) and (\ref{eq:3.4}). Despite the difference in the forcing profiles shown in figure \ref{fprofile}, the streamwise wavenumber spectra of the wall-normal velocity reveal that the energetic vortical structures at $\lambda_x \simeq 1-2h$ in the outer region emerge in all the three cases (figures \ref{viscspectra}$b$,$c$,$d$). This clearly indicates that the streak instability is a very robust physical process that depends very little on the type of the forcing, as long as the forcing is kept to generate an amplified streak in the outer region.

\section{Concluding remarks}\label{sec:Con}
Thus far, we have examined the instability of an `amplified' streaky motion artificially driven by the optimal forcing computed in \cite{Hwang2010b}. As the forcing amplitude is gradually increased, a streamwise-uniform streak, reminiscent of VLSM, is successfully introduced. We have shown that an energetic cross-streamwise velocity structure emerges at the streamwise length scale $\lambda_x/h =1-2$, and this structure produces all statistical features of the LSM remarkably. All diagnosis tools employed (the cross-streamwise turbulence statistics, DMD analysis and energy-production analysis) firmly indicate that the origin of this structure is a sinuous-mode streak instability. Finally, it has been proposed that the streak instability mechanism determines the streamwise length scales of the LSM and the VLSM.

It is evident that the present numerical experiment directly supports the existence of the self-sustaining process in the outer region \citep{Hwang2010d,Hwang2015,Hwang2016}. In particular, it should be noted that the DMD structure given in figure \ref{InstvdDMD} ($c$) is remarkably similar to the invariant solution in \cite{Hwang2016b}. However, the most important contribution of the present study is that it provides direct evidence of the streak instability process, the existence of which has only been speculated for many years. The present study also unveils the importance of nonlinear processes in understanding the mechanism of the streamwise length-scale determination, and this may play a crucial role in developing low-order models for coherent structures at high Reynolds numbers. Finally, it should be mentioned that the present approach can be extended to the logarithmic region, where the self-sustaining process has been recently shown to emerge in a hierarchical manner \cite[]{Hwang2015,Hwang2016}. Our efforts are focused in this direction, and we believe that this will unveil the in detail formation mechanism of the self-similar vortex clusters in the logarithmic region \cite[]{delAlamo2006}.

\section*{Acknowledgements}
\begin{acknowledgments}
This work was supported by the Engineering and Physical Sciences Research Council (EPSRC) in the UK (EP/N019342/1). We also thank the anonymous reviewers for a number of useful suggestions.
\end{acknowledgments}

\appendix

\section{Dynamic Mode Decomposition}\label{DMDapp}
As stated in \S \ref{ssec:DMD}, the parameters for the DMD need to be chosen very carefully. Hereafter, we discuss how we have chosen the three main parameters of the DMD: i.e. the sampling time-interval $\Delta t$, the number of snapshots $N$, and the number of POD modes $r$.

\begin{figure}
 \centerline{\includegraphics[width=.59\textwidth]{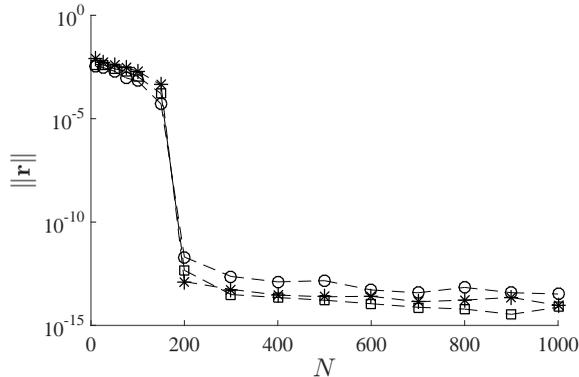}}
  \caption{The norm of the residual vector with the number of snapshots $N$ (S960): $*$, $A^+_s=3.42$; {\tiny$\square$}, $A^+_s=4.04$; {\tiny$\lozenge$}, $A^+_s=4.76$.}
\label{fig:app1}
\end{figure}

\subsection{Sampling time interval $\Delta t$}
The theoretical upper bound for the sampling time step is represented by the Nyquist criterion, which applies to the frequency of the process we are interested in (i.e. streak instability). There is also a lower bound, since a too high sampling rate could result in this feature being perceived as quasi-steady \citep{Schmid2010}. The typical time scale of the self-sustaining process in the outer layer is $T_{ssp} u_\tau/h \simeq 3$ and the time scale of the streak instability is even shorter \cite[]{Hwang2016}. To provide a good resolution for this time scale, four different snapshot time intervals have been tested with S960: $\Delta t u_\tau/h = 0.009, ~0.018,~0.027,~0.036$. The two smaller values yield eigenvalues of $\mathbf{{S}}$ lying almost perfectly on the unit circle $|\mu|=1$ (see figure \ref{fig:unitc}), and the smallest value has been used for the DMD analysis in the present study.

\subsection{Total number of snapshots $N$}
The minimum number of snapshots $N$ is determined as follows. We first assume that the $N$-th snapshot is given by a linear combination of all previous $N-1$ snapshots with a residual vector $\mathbf{r}$: i.e.
\begin{equation}\label{residual_1}
\psi_N = \mathbf{\Psi_0} \mathbf{a} + \mathbf{r},
\end{equation}
where $\mathbf{a} = [a_1~ a_2~ ... a_{N-1}]^T$ is a column vector for the coefficients of the linear combination. If the norm of $\mathbf{r}$ approaches zero for sufficiently large $N$, the number of snapshots may be considered large enough to cover any flow field using the snapshots. The minimum residual $\mathbf{r}$ is computed with the economy-size QR decomposition of the snapshots $\mathbf{\Psi_0} = \mathbf{Q} \mathbf{R}$, and the coefficients of the linear combination are given by
\begin{equation}\label{residual_2}
\mathbf{a} = \mathbf{R}^{-1} \mathbf{Q}^H \psi_N.
\end{equation}
The residual and its 2-norm are then found with (\ref{residual_1}) and (\ref{residual_2}).

Fig. \ref{fig:app1} shows the dependence of the 2-norm of \textbf{r} on the number of snapshots. The test was carried out with $\Delta t u_\tau/h = 0.009$ using the S960 simulations with three different forcing amplitudes. At $N\simeq200$, the 2-norm of the residual vector drops drastically, and falls below $O(10^{-10})$. However, further increase in the number of snapshots $N$ does not decrease the residual vector significantly, and it remains at $O(10^{-15})$ even for $N=1000$. The number of snapshots chosen in the present study is $N=1000$. We note that this number also ensures that the total time interval of the snapshots is $T u_\tau/h \simeq 9$, which would be six to nine times longer than the streak meandering time scale observed in \cite{Hwang2016}.

\begin{figure}
 \centerline{\includegraphics[width=.55\textwidth]{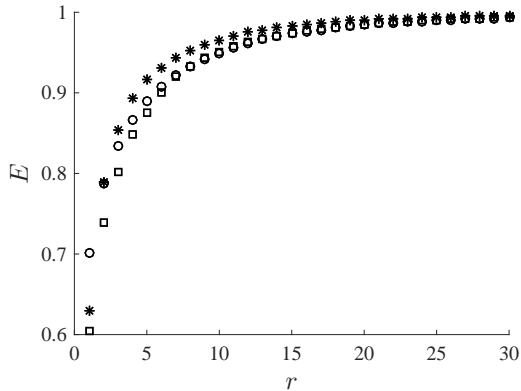}}
  \caption{Energy content of the first $r$ POD modes (S960):  $*$,$~A^+_s=3.42$; {\tiny$\square$}, $A^+_s=4.04$; {\tiny$\lozenge$}, $A^+_s=4.76$.}
\label{fig:app2}
\end{figure}

\subsection{Number of POD modes \textit{r}}
The choice of the number of POD modes to be included has been based on the amount of energy contained in the first \textit{r} POD modes. Here, $N=1000$ and $\Delta t u_\tau/h =0.009$, and the calculations were performed with the $S960$ simulations. If the matrix of snapshots is decomposed as $\boldsymbol{\Psi}_0=\mathbf{U}\Sigma\mathbf{V}^H$, the relative energy content up to the $r^{\mathrm{th}}$ mode is computed by:
\begin{equation}
E=\frac{{\sum\limits^r_{i=1}\Sigma_{ii}}}{trace(\Sigma)}.
\end{equation}
Fig. \ref{fig:app2} reports the dependence of $E$ on $r$. With only the first twenty-five modes, more than 95\% of the total energy is recovered. Given this feature, it is evident that the use of all the POD modes is not necessary. Furthermore, appropriate truncation of the number of POD modes would help for eliminating noise associated with background turbulence, as higher-order POD modes are expected to contain such noise. Therefore, in the present study, thirty POD modes are used to ensure to contain all the energetic features in our DMD analysis, while appropriately removing the effect of noisy small-scale background in the snapshots.

\bibliography{references}
\bibliographystyle{jfm}

\end{document}